\documentclass[
  aps,
  pre,
  reprint,
  floatfix,
  superscriptaddress,
  amsmath,
  amssymb
]{revtex4-2}

\usepackage{bm}
\usepackage{graphicx}
\usepackage{subfigure}
\usepackage{upgreek}
\usepackage{hyperref}
\hypersetup{hidelinks}
\graphicspath{{./}}

\begin{document}

\title{Rigorously justified local time stepping in the unified gas-kinetic wave-particle method for steady multiscale flow simulation}

\author{Wenzhi Guo}
\email{wguoal@connect.ust.hk}
\affiliation{Department of Mathematics, Hong Kong University of Science and Technology, Hong Kong, China}

\author{Junzhe Cao}
\email{jcaobb@connect.ust.hk}
\affiliation{Department of Mathematics, Hong Kong University of Science and Technology, Hong Kong, China}

\author{Wenpei Long}
\email{wlongab@connect.ust.hk}
\affiliation{Department of Mathematics, Hong Kong University of Science and Technology, Hong Kong, China}

\author{Kun Xu}
\email{Corresponding author: makxu@ust.hk}
\affiliation{Department of Mathematics, Hong Kong University of Science and Technology, Hong Kong, China}
\affiliation{Department of Mechanical and Aerospace Engineering, Hong Kong University of Science and Technology, Hong Kong, China}
\affiliation{HKUST Shenzhen Research Institute, Shenzhen, 518057, China}

\begin{abstract}
Local time stepping (LTS) can accelerate convergence to steady states in kinetic simulations with large variations in the local time steps across the computational domain. When neighboring cells advance with unequal time steps, the time-averaged particle flux must be balanced across their common interface. For particle-based or hybrid wave-particle methods under finite volume method (FVM) framework, we rigorously establish a sufficient condition for time-averaged interfacial particle-flux balance: fixed positive local time steps together with proportional particle-mass scaling. When a particle crosses from cell $L$ to cell $R$, its mass is scaled by $\Delta t_R/\Delta t_L$. In the unified gas-kinetic wave-particle (UGKWP) implementation, the same ratio is applied to the remaining free-transport time of the crossing particle. LTS also affects the wave-particle decomposition and the time integration of the wave fluxes in UGKWP. The wave-particle decomposition in cell \(i\) is determined by the local ratio \(\Delta t_i/\tau_i\), which better reflects the local relation between the observation scale and relaxation time for multiscale cases. The equilibrium and analytic free transport wave fluxes are integrated over and normalized by the corresponding cell-side time steps to obtain the interfacial time-averaged wave fluxes. The UGKWP-LTS method is used to simulate the hypersonic flow past a cylinder at $\mathrm{Kn}=0.01$ and $0.1$, and a flat plate at $\mathrm{Kn}=0.0169$. In all three cases, the surface quantities obtained with UGKWP-LTS agree well with the reference data. Relative to global time stepping (GTS), UGKWP-LTS achieves step-count speedups of $6.6\times$, $3.8\times$, and $20\times$ for the three cases, respectively. The corresponding wall-clock speedups are $7.1\times$, $4.5\times$, and approximately $21\times$.
\end{abstract}

\maketitle

\section{Introduction}

\label{sec:introduction}
Multiscale gas flows encompass continuum and rarefied regimes, in which molecular free transport and collisions interact across different spatial and temporal scales. For multiscale gas flow modeling, Xu proposed the direct modeling framework based on the integral solution of the kinetic equation along a characteristic line~\cite{dm1,dm2}. The finite $\Delta t$ in this integral solution couples molecular free transport and collisions over the observation scale, thereby yielding a space-time-coupled multiscale evolution. Within this framework, the unified gas-kinetic wave-particle (UGKWP) method combines a deterministic wave representation with stochastic particles for multiscale flow evolution~\cite{ugkwp1,ugkwp2}. The ratio $\Delta t/\tau$ governs the adaptive wave-particle decomposition across flow regimes. As $\Delta t/\tau$ increases, the wave representation becomes dominant and UGKWP recovers the gas-kinetic scheme (GKS)~\cite{gks2001}; as $\Delta t/\tau$ decreases, stochastic particles become increasingly important. Unlike the unified gas-kinetic scheme (UGKS), which evolves the gas distribution function on a discrete velocity space~\cite{ugks}, UGKWP represents the nonequilibrium part of the distribution function with stochastic particles. Recently, UGKWP has been applied to diatomic molecular flows with internal nonequilibrium~\cite{ugkwp5,wp-vib}, dilute gas-particle multiphase flow~\cite{wp-phase}, radiative transport~\cite{ugkwp3,wp-rad}, plasma transport~\cite{ugkwp4,wp-plasma}, and nonequilibrium turbulent flows~\cite{wp-turb1,wp-turb2}. Related theoretical developments include the kinetic representation of UGKWP~\cite{wp-6}, a unified gas-kinetic framework from the Boltzmann to Navier--Stokes scales~\cite{ugkf}, nonequilibrium gas modeling beyond a single relaxation time description~\cite{xu-tau}, and a wave-particle decomposition of the governing equation based on a local kinetic horizon~\cite{wpd2026}. Simplified wave-particle formulations, including the simplified unified wave-particle (SUWP) method and the simple hydrodynamic-particle method (SHPM), have also been developed~\cite{suwp,shpm}.

Efficient steady-state computation is important in large-scale aerodynamic simulations, and implicit solvers on GPUs have been developed to improve computational efficiency~\cite{yang2025mcgs}. In steady hypersonic flow computations, the first-layer mesh height near the wall is usually very small. The CFL condition therefore requires a small local time step in these cells. Under global time stepping (GTS), the smallest time step \(\Delta t_G\) is used throughout the computational domain. For a steady flow with a unique solution, a consistent local time stepping (LTS) formulation changes the convergence process rather than the converged state. LTS can accelerate convergence by using a locally determined time step in each cell. For UGKWP, however, LTS requires both the deterministic wave flux and stochastic particles crossing an interface to satisfy the same time-averaged flux requirement, especially between neighboring cells with different observation scales, where the local ratio $\Delta t_i/\tau_i$ determines the wave-particle decomposition in each cell.

LTS has been studied in direct simulation Monte Carlo (DSMC) and related particle methods. Within the DSMC framework~\cite{dsmc}, variable time steps, variable particle weights, and grid manipulation were examined as strategies for improving particle resolution~\cite{boyd2000}. Wu et al. formulated a particle-flux relation for a single interface crossing, requiring the particle weight to scale proportionally with the local time step, and rescaling the remaining transport time according to the interface time-step ratio~\cite{wu2004vts}. Later DSMC implementations applied LTS to parallel micro/nano-cavity simulations and GPU-accelerated cut-cell calculations~\cite{roohi2013,wu2014}. An adaptive DSMC procedure updated the local time steps and particle weights as the simulation progressed toward steady state~\cite{boyd2015jcp}. Related developments in particle-based hypersonic simulation were reviewed in Ref.~\cite{boyd2015pas}. Spatiotemporal adaptation with LTS was introduced into the unified stochastic particle method~\cite{usp2024}. Recently, an LTS study for SUWP modified the hybrid framework to address the weight mismatch between the interfacial flux and cell-centered particles~\cite{suwp2026}. These studies concentrate on one individual particle within one time step. Practical implementations additionally employ method-specific controls. For example, the adaptive DSMC procedure limits the time-step update magnitude and smooths the local time steps~\cite{boyd2015jcp}. However, no sufficient condition has been established for time-averaged interfacial particle-flux balance over multiple time steps.

For particle-based or hybrid wave-particle methods under finite volume method (FVM) framework, we provide the first rigorous proof that two conditions are sufficient for time-averaged interfacial particle-flux balance: fixed positive local time steps and proportional scaling of the mass of particles crossing an interface. No additional restriction is imposed on the time-step ratio or on the spatial variation of the fixed local time steps. In UGKWP-LTS, the interfacial flux consists of the collided equilibrium wave flux, the free transport equilibrium wave flux, and the free transport flux of stochastic particles. For each wave flux, the time-averaged fluxes evaluated from the two sides of the interface are assembled into a single time-averaged interfacial flux. For the update of each adjacent cell, this flux is multiplied by that cell's local time step to recover the corresponding time-integrated flux. For stochastic particles, the mass of a particle crossing from cell \(L\) to cell \(R\) is scaled by \(\Delta t_R/\Delta t_L\) to satisfy the time-averaged particle-flux balance. The remaining free transport time of the crossing particle is updated using the same ratio. The wave-particle decomposition in cell \(i\) is determined by the local ratio \(\Delta t_i/\tau_i\).

This paper is organized as follows. Section~\ref{sec:gas-kinetic-theory} introduces the gas-kinetic theory and model. Section~\ref{sec:ugkwp-method} presents the UGKWP method and its wave-particle evolution procedure. Section~\ref{sec:conservation-analysis} presents the LTS formulation, establishes the sufficient condition for time-averaged interfacial particle-flux balance, and describes its implementation in UGKWP. Section~\ref{sec:verification} presents numerical verification using hypersonic flow past a cylinder and a flat plate. Section~\ref{sec:conclusions} gives the conclusions.

\section{Gas-kinetic theory}
\label{sec:gas-kinetic-theory}
In gas-kinetic theory, the gas state is described by the gas distribution function \(f(\boldsymbol{x},\boldsymbol{u},\boldsymbol{\xi},t)\), where \(\boldsymbol{x}\), \(\boldsymbol{u}\), \(\boldsymbol{\xi}\), and \(t\) denote the position, molecular velocity, equivalent velocity representing the internal energy, and time, respectively. Here, \(\boldsymbol{\xi}\in\mathbb{R}^{D}\), and \(D\) is the number of internal degrees of freedom. The local equilibrium state $g$ is given by the Maxwellian distribution
\begin{equation}
\label{eq:maxwellian}
g=\rho\left(\frac{1}{2\pi RT}\right)^{\frac{D+3}{2}}
\exp\left[-\frac{|\boldsymbol{c}|^2+|\boldsymbol{\xi}|^2}{2RT}\right],
\end{equation}
where $\rho$, $\boldsymbol{U}$, and $T$ denote the density, macroscopic velocity, and temperature, respectively, $R$ is the specific gas constant, and $\boldsymbol{c}=\boldsymbol{u}-\boldsymbol{U}$ is the peculiar velocity. The evolution of the gas distribution function is governed by the Boltzmann equation,
\begin{equation}
\label{eq:boltzmann}
\frac{\partial f}{\partial t}+\boldsymbol{u}\cdot\nabla_{\boldsymbol{x}}f=Q(f,f),
\end{equation}
where $Q(f,f)$ is the Boltzmann collision operator. The BGK model~\cite{bgk} approximates the collision operator by a single relaxation toward the local Maxwellian:
\begin{equation}
\label{eq:bgk}
\frac{\partial f}{\partial t}+\boldsymbol{u}\cdot\nabla_{\boldsymbol{x}}f=\frac{g-f}{\tau}.
\end{equation}
The relaxation time $\tau$ characterizes the time scale of relaxation toward equilibrium. Its definition is $\tau=\mu/p$, where $\mu$ is the dynamic viscosity and $p=\rho RT$ is the pressure. The Shakhov model~\cite{shakhov} introduces a Prandtl-number correction through the modified distribution $g^S$:
\begin{equation}
\label{eq:shakhov}
\begin{gathered}
\frac{\partial f}{\partial t} +\boldsymbol{u}\cdot\nabla_{\boldsymbol{x}}f =\frac{g^S-f}{\tau},\\
g^S = g\left[1+(1-\mathrm{Pr}) \frac{\boldsymbol{c}\cdot\boldsymbol{Q}}{5pRT} \left(\frac{|\boldsymbol{c}|^2}{RT}-5\right)\right],
\end{gathered}
\end{equation}
where $\mathrm{Pr}$ is the Prandtl number. The heat-flux vector $\boldsymbol{Q}$ is defined as
\begin{equation}
\label{eq:heat-flux}
\boldsymbol{Q} = \int_{\mathbb{R}^{D}}\int_{\mathbb{R}^{3}} \boldsymbol{c}
\frac{|\boldsymbol{c}|^2+|\boldsymbol{\xi}|^2}{2} f\,\mathrm{d}\boldsymbol{u}\,\mathrm{d}\boldsymbol{\xi}.
\end{equation}
When $\mathrm{Pr}=1$, the Shakhov model recovers the BGK model. The collision-invariant vector is defined as
\begin{equation}
\label{eq:kinetic-moments}
\boldsymbol{\Psi}=\left(1,\boldsymbol{u},\frac{1}{2}\left(|\boldsymbol{u}|^2+|\boldsymbol{\xi}|^2\right)\right)^T.
\end{equation}
The corresponding macroscopic conservative variables are
\begin{equation}
\label{eq:conservative-moments}
\boldsymbol{W}=(\rho,\rho\boldsymbol{U},\rho E)^T
=\int_{\mathbb{R}^{D}}\int_{\mathbb{R}^{3}}\boldsymbol{\Psi}f\,\mathrm{d}\boldsymbol{u}\,\mathrm{d}\boldsymbol{\xi},
\end{equation}
where $E$ is the total energy per unit mass. The collision model must preserve mass, momentum, and total energy. This requirement is expressed by the compatibility condition
\begin{equation}
\label{eq:compatibility}
\int_{\mathbb{R}^{D}}\int_{\mathbb{R}^{3}}\boldsymbol{\Psi}\left(g^S-f\right)\,\mathrm{d}\boldsymbol{u}\,\mathrm{d}\boldsymbol{\xi}=\boldsymbol{0},
\end{equation}
which ensures that $g^S$ and $f$ have the same conservative moments.

\section{Unified gas-kinetic wave-particle method}
\label{sec:ugkwp-method}
Based on the preceding kinetic model, UGKWP evaluates the time-integrated numerical flux by combining deterministic wave evolution with stochastic-particle transport. The UGKWP method is constructed under FVM framework. For cell $i$ at step $n$, the governing equations for the cell-averaged macroscopic conservative variables $\boldsymbol{W}_i^n$ and gas distribution function $f_i^n$ are
\begin{equation}
\label{eq:conservation}
\begin{gathered}
\boldsymbol{W}^{n+1}_i = \boldsymbol{W}^{n}_i - \frac{1}{\Omega_i} \sum_{j\in\mathcal{M}(i)} \boldsymbol{F}_j s_j,\\
f^{n+1}_i = f^n_i - \frac{1}{\Omega_i} \sum_{j\in\mathcal{M}(i)} \int_0^{\Delta t}
\boldsymbol{u}\cdot\boldsymbol{n}_j f_j(t)s_j\,\mathrm{d}t + \int_0^{\Delta t}\mathcal{J}_i\,\mathrm{d}t,
\end{gathered}
\end{equation}
where \(\mathcal{M}(i)\) is the set of interfaces of cell \(i\), \(s_j\) is the length or area of interface \(j\), \(\boldsymbol{n}_j\) is the outward unit normal vector of cell \(i\) at interface \(j\), and \(\Omega_i\) is the cell area or volume. Moreover, \(f_j(t)\) is the time-dependent gas distribution function at interface \(j\), and \(\mathcal{J}_i\) denotes the collision term in cell \(i\). The time-integrated macroscopic flux through interface $j$ is
\begin{equation}
\label{eq:macrof}
\boldsymbol{F}_j = \int_{\mathbb{R}^{D}}\int_{\mathbb{R}^{3}}\int^{\Delta t}_0\boldsymbol{u}\cdot\boldsymbol{n}_j f_j(t)\boldsymbol{\Psi}\,\mathrm{d}t\,\mathrm{d}\boldsymbol{u}\,\mathrm{d}\boldsymbol{\xi}.
\end{equation}

To evaluate the time-integrated flux in Eq.~\eqref{eq:macrof}, the time-dependent gas distribution function $f_j(t)$ at interface $j$ is constructed from the integral solution of the kinetic model along the characteristic line. Taking the interface center as the spatial origin and the beginning of the current time step as $t=0$, the integral solution is
\begin{equation}
\label{eq:integral}
\begin{aligned}
f(\boldsymbol{0},t) ={}& \frac{1}{\tau}\int_0^t
g^S\left[-\boldsymbol{u}(t-\tilde{t}),\tilde{t}\right]
e^{-(t-\tilde{t})/\tau}\,\mathrm{d}\tilde{t} \\
&+e^{-t/\tau}f(-\boldsymbol{u}t,0).
\end{aligned}
\end{equation}
For the second-order flux construction, $g^S$ is expanded in space and time about $(\boldsymbol{0},0)$, while the initial distribution is reconstructed in space at $t=0$ as
\begin{equation}
\label{eq:second-order-expansion}
\begin{aligned}
g^S(\boldsymbol{x},t) &= g^S_0+\frac{\partial g^S}{\partial\boldsymbol{x}}\cdot\boldsymbol{x}+\frac{\partial g^S}{\partial t}t,\\
f(\boldsymbol{x},0) &= f_0+\frac{\partial f}{\partial\boldsymbol{x}}\cdot\boldsymbol{x}.
\end{aligned}
\end{equation}
Substituting Eq.~\eqref{eq:second-order-expansion} into Eq.~\eqref{eq:integral} gives the time-dependent interface distribution
\begin{equation}
\label{eq:time-dependent-distribution}
\begin{aligned}
f(\boldsymbol{0},t)={}&\varepsilon_a g^S_0
+ \varepsilon_b \frac{\partial g^S}{\partial\boldsymbol{x}}\cdot\boldsymbol{u} + \varepsilon_c \frac{\partial g^S}{\partial t}
+ \varepsilon_d f_0 + \varepsilon_e \frac{\partial f}{\partial\boldsymbol{x}}\cdot\boldsymbol{u},\\
\varepsilon_a={}&1-e^{-t/\tau},\\
\varepsilon_b={}&te^{-t/\tau}-\tau(1-e^{-t/\tau}),\\
\varepsilon_c={}&t-\tau(1-e^{-t/\tau}),\\
\varepsilon_d={}&e^{-t/\tau},\\
\varepsilon_e={}&-te^{-t/\tau}.
\end{aligned}
\end{equation}
Taking moments of Eq.~\eqref{eq:time-dependent-distribution} and integrating over \([0,\Delta t]\) gives the time-integrated flux decomposition
\begin{equation}
\label{eq:integralb1}
\boldsymbol{F}_j=\boldsymbol{F}^{eq}_j+\boldsymbol{F}^{fr}_j,
\end{equation}
where the equilibrium and free transport terms are
\begin{equation}
\label{eq:integralb2}
\begin{aligned}
\boldsymbol{F}^{eq}_j ={}& \int_{\mathbb{R}^D}\int_{\mathbb{R}^3}\boldsymbol{\Psi}
\left(\delta_ag^S_0+\delta_b\frac{\partial g^S}{\partial \boldsymbol{x}}\cdot\boldsymbol{u}
+\delta_c\frac{\partial g^S}{\partial t}\right) \\
&\quad\left(\boldsymbol{u}\cdot\boldsymbol{n}_j\right)
\,\mathrm{d}\boldsymbol{u}\,\mathrm{d}\boldsymbol{\xi},\\
\boldsymbol{F}^{fr}_j ={}& \int_{\mathbb{R}^D}\int_{\mathbb{R}^3}\boldsymbol{\Psi}
\left(\delta_df_0+\delta_e\frac{\partial f}{\partial \boldsymbol{x}}\cdot\boldsymbol{u}\right) \\
&\quad\left(\boldsymbol{u}\cdot\boldsymbol{n}_j\right)
\,\mathrm{d}\boldsymbol{u}\,\mathrm{d}\boldsymbol{\xi}.
\end{aligned}
\end{equation}
The time-integration coefficients are
\begin{equation}
\label{eq:time-integration-coefficients}
\begin{aligned}
\delta_a &= \Delta t-\tau\left(1-e^{-\Delta t/\tau}\right),\\
\delta_b &= 2\tau^2\left(1-e^{-\Delta t/\tau}\right)-\tau\Delta t-\tau\Delta t e^{-\Delta t/\tau},\\
\delta_c &= \frac{\Delta t^2}{2}-\tau\Delta t+\tau^2\left(1-e^{-\Delta t/\tau}\right),\\
\delta_d &= \tau\left(1-e^{-\Delta t/\tau}\right),\\
\delta_e &= \tau\Delta t e^{-\Delta t/\tau}-\tau^2\left(1-e^{-\Delta t/\tau}\right).
\end{aligned}
\end{equation}
In UGKWP, $\boldsymbol{F}^{eq}_j$ denotes the equilibrium wave flux evaluated analytically. The free transport flux is decomposed as
\[
    \boldsymbol{F}^{fr}_j=\boldsymbol{F}^{fr,wave}_j+\boldsymbol{F}^{fr,p}_j,
\]
where $\boldsymbol{F}^{fr,wave}_j$ is the analytic flux from the non-sampled hydrodynamic wave and $\boldsymbol{F}^{fr,p}_j$ is evaluated from stochastic particles crossing the interface. Consequently, the UGKWP flux consists of $\boldsymbol{F}^{eq}_j$, $\boldsymbol{F}^{fr,wave}_j$ and $\boldsymbol{F}^{fr,p}_j$.

\subsection{Wave evolution}
\label{sec:wave-evolution}
For the equilibrium wave flux $\boldsymbol{F}^{eq}_j$, the interface equilibrium state and its spatial and temporal derivatives are constructed following the GKS~\cite{gks1998,gks2001}. For the Shakhov model, $g_0^S$, $\partial g^S/\partial\boldsymbol{x}$ and $\partial g^S/\partial t$ are evaluated using the Maxwellian distribution $g_0$ and its corresponding derivatives $\partial g/\partial\boldsymbol{x}$ and $\partial g/\partial t$~\cite{ugkwp5}. The interface conservative state $\boldsymbol{W}_0$ is determined from the half-range moments of the reconstructed left and right Maxwellian distributions:
\begin{equation}
\label{eq:interface-equilibrium}
\boldsymbol{W}_0
=\int\boldsymbol{\Psi}g_0\,\mathrm{d}\Xi
=\int_{\boldsymbol{u}\cdot\boldsymbol{n}_j>0}
\boldsymbol{\Psi}g_L\,\mathrm{d}\Xi
+\int_{\boldsymbol{u}\cdot\boldsymbol{n}_j<0}
\boldsymbol{\Psi}g_R\,\mathrm{d}\Xi,
\end{equation}
where $g_L$ and $g_R$ are the Maxwellian distributions constructed from the reconstructed macroscopic states on the left and right sides of the interface, respectively, and $\mathrm{d}\Xi=\mathrm{d}\boldsymbol{u}\,\mathrm{d}\boldsymbol{\xi}$. The spatial and temporal derivatives of the Maxwellian distribution are written as
\begin{equation}
\label{eq:wave-derivatives}
\frac{\partial g}{\partial\boldsymbol{x}}=g\boldsymbol{a},
\qquad
\frac{\partial g}{\partial t}=gA.
\end{equation}
The spatial derivative coefficient $\boldsymbol{a}$ is expanded as
\begin{equation}
\label{eq:wave-spatial-coefficient}
\boldsymbol{a}
=\boldsymbol{a}_0+\boldsymbol{a}_1 u_1 + \boldsymbol{a}_2 u_2 + \boldsymbol{a}_3 u_3
+ \boldsymbol{a}_4\frac{|\boldsymbol{u}|^2+|\boldsymbol{\xi}|^2}{2},
\end{equation}
and its coefficients are determined from
\begin{equation}
\label{eq:wave-spatial-moment}
\int\boldsymbol{\Psi}g\boldsymbol{a}\,\mathrm{d}\Xi
=\frac{\partial\boldsymbol{W}_0}{\partial\boldsymbol{x}}.
\end{equation}
With $\Lambda=1/(2RT)$, the Prandtl correction~\cite{may2007} is applied by replacing $\partial\Lambda/\partial\boldsymbol{x}$ with $(\partial\Lambda/\partial\boldsymbol{x})/\mathrm{Pr}$ in the evaluation of $\boldsymbol{a}$.
\begin{equation}
\label{eq:wave-temporal-expansion}
A = A_0 + A_1 u_1 + A_2 u_2 + A_3 u_3 + A_4\frac{|\boldsymbol{u}|^2+|\boldsymbol{\xi}|^2}{2},
\end{equation}
where its coefficients are determined from the compatibility relation
\begin{equation}
\label{eq:wave-temporal-coefficient}
\int\boldsymbol{\Psi}gA\,\mathrm{d}\Xi
=-\int\boldsymbol{\Psi}g
\left(\boldsymbol{a}\cdot\boldsymbol{u}\right)\,\mathrm{d}\Xi.
\end{equation}

When the relaxation time varies sharply across an interface, the time-integration coefficients in the equilibrium wave flux are evaluated separately using the cell-side relaxation times, so that the coefficients used in the wave flux are consistent with those associated with the cell-side particle evolution~\cite{awp2}. For GTS, the left and right coefficients are evaluated from Eq.~\eqref{eq:time-integration-coefficients} using $(\Delta t_G,\tau_L)$ and $(\Delta t_G,\tau_R)$, respectively. The equilibrium wave flux at interface $j$ is
\begin{equation}
\label{eq:feq}
\begin{aligned}
\boldsymbol{F}^{eq}_{j}={}&
\int_{\mathbb{R}^{D}} \int_{\boldsymbol{u}\cdot\boldsymbol{n}_{j}>0} \boldsymbol{\Psi}
\left(
\delta_{a,L}g_0
+\delta_{b,L}\frac{\partial g}{\partial\boldsymbol{x}}\cdot\boldsymbol{u}
+\delta_{c,L}\frac{\partial g}{\partial t}
\right)
\\
&
\qquad\left(\boldsymbol{u}\cdot\boldsymbol{n}_{j}\right)
\,\mathrm{d}\boldsymbol{u}\,\mathrm{d}\boldsymbol{\xi}
\\
&+
\int_{\mathbb{R}^{D}} \int_{\boldsymbol{u}\cdot\boldsymbol{n}_{j}<0} \boldsymbol{\Psi}
\left(
\delta_{a,R}g_0
+\delta_{b,R}\frac{\partial g}{\partial\boldsymbol{x}}\cdot\boldsymbol{u}
+\delta_{c,R}\frac{\partial g}{\partial t}
\right)
\\
&
\qquad\left(\boldsymbol{u}\cdot\boldsymbol{n}_{j}\right)
\,\mathrm{d}\boldsymbol{u}\,\mathrm{d}\boldsymbol{\xi}.
\end{aligned}
\end{equation}

\subsection{Particle transport}
\label{sec:particle-transport}
To construct the analytic free transport wave flux and sample stochastic particles, the cell-averaged conservative variables are decomposed into hydrodynamic-wave and particle components,
\begin{equation}
\label{eq:wave-particle-decomposition}
\boldsymbol{W}_i=\boldsymbol{W}_i^h+\boldsymbol{W}_i^p.
\end{equation}
At the beginning of each time step, $\boldsymbol{W}_i^p$ is evaluated from the particles retained in cell $i$ from the preceding step, and the hydrodynamic-wave state is obtained as
\[
    \boldsymbol{W}_i^h=\boldsymbol{W}_i-\boldsymbol{W}_i^p.
\]
A simulation particle $P_k(m_k,\boldsymbol{x}_k,\boldsymbol{u}_k,\boldsymbol{\xi}_k)$ is characterized by its mass $m_k$, position $\boldsymbol{x}_k$, molecular velocity $\boldsymbol{u}_k$, and equivalent velocity $\boldsymbol{\xi}_k$. The corresponding collision-invariant vector is
\[
\boldsymbol{\Psi}_k = \left(1,\boldsymbol{u}_k, \frac{1}{2} \left( |\boldsymbol{u}_k|^2+|\boldsymbol{\xi}_k|^2 \right) \right)^T.
\]
 At the beginning of each GTS step, the particles to be transported consist of particles retained from the preceding step and collisionless particles newly sampled from the hydrodynamic-wave state. Since all cells use the common time step $\Delta t_G$, the conservative state represented by the newly sampled collisionless particles is $\boldsymbol{W}_i^{hp}=\boldsymbol{W}_i^h e^{-\Delta t_G/\tau_i}$, whose density component is
\begin{equation}
\label{eq:wave-resampling}
\rho_i^{hp} = \rho_i^h e^{-\Delta t_G/\tau_i}.
\end{equation}
For $\rho_i^{hp}>0$, the number and mass of the newly sampled particles are
\begin{equation}
\label{eq:particle-sampling}
N_i^{hp} = \left\lceil \frac{\rho_i^{hp}}{\rho_i}N_{\mathrm{ref}} \right\rceil,
\qquad
m_k = \frac{\rho_i^{hp}\Omega_i}{N_i^{hp}},
\end{equation}
where \(\rho_i\) is the total density in cell \(i\), and $N_{\mathrm{ref}}$ is the reference particle number per cell. Particle positions are sampled uniformly within the cell. Molecular velocities are sampled from the local Shakhov distribution $g^S$ using the acceptance-rejection (A-R) method, and $\boldsymbol{\xi}_k$ is sampled according to the cell temperature $T_i$ and the number of internal degrees of freedom $D$. By construction, the newly sampled particles represent the collisionless part of the hydrodynamic wave over the current observation scale and are assigned $t_{f,k}=\Delta t_G$.

For a particle retained from the preceding step and located in cell $i$ at the beginning of the current step, a uniform random number $\epsilon_k\in(0,1)$ is generated. According to Eq.~\eqref{eq:integral}, its free-transport time is determined by
\begin{equation}
\label{eq:mic1}
t_{f,k} = \min\left[-\tau_i\ln(\epsilon_k),\Delta t_G\right].
\end{equation}
Both retained and newly sampled particles are transported according to
\begin{equation}
\label{eq:mic2}
\boldsymbol{x}_k \leftarrow \boldsymbol{x}_k+\boldsymbol{u}_kt_{f,k}.
\end{equation}
Cell-interface crossings along the transport path are detected and processed during particle transport.

After particle transport and before collisional-particle removal, the net change in the conservative variables represented by stochastic particles is evaluated directly from the pre-transport and post-transport particle sets:
\begin{equation}
\label{eq:mic3}
\boldsymbol{W}_i^{fr,p} = \boldsymbol{W}_i^{p,+} - \boldsymbol{W}_i^{p,-},
\end{equation}
where
\begin{equation}
\label{eq:mic2b}
\boldsymbol{W}_i^{p,\pm} = \frac{1}{\Omega_i} \sum_{k\in\mathcal{N}^{\pm}(i)} m_k\boldsymbol{\Psi}_k.
\end{equation}
Here $\mathcal{N}^{-}(i)$ and $\mathcal{N}^{+}(i)$ denote the particle sets located in cell $i$ immediately before and after free transport, respectively. Particles with $t_{f,k}<\Delta t_G$ are then removed and represented by the hydrodynamic wave, whereas particles with $t_{f,k}=\Delta t_G$ are retained for the next time step.

The portion $\boldsymbol{W}_i^h-\boldsymbol{W}_i^{hp}$ is not represented by stochastic particles. Its analytic free transport wave flux is obtained by subtracting the flux already represented by the newly sampled collisionless particles from the full free transport flux associated with the hydrodynamic-wave state. Following the side-dependent coefficient evaluation in Ref.~\cite{awp2}, the left and right coefficients are evaluated from Eq.~\eqref{eq:time-integration-coefficients} using $(\Delta t_G,\tau_L)$ and $(\Delta t_G,\tau_R)$, respectively. The analytic free transport wave flux is
\begin{equation}
\label{eq:free-wave-flux}
\begin{aligned}
\boldsymbol{F}^{fr,wave}_j={}&
\int_{\mathbb{R}^{D}} \int_{\boldsymbol{u}\cdot\boldsymbol{n}_j>0} \boldsymbol{\Psi}
\Bigg[
\left( \delta_{d,L}-\Delta t_Ge^{-\Delta t_G/\tau_L} \right)g_0^h
\\
&\quad+
\left( \delta_{e,L} + \frac{\Delta t_G^2}{2}e^{-\Delta t_G/\tau_L} \right)
\frac{\partial g^h}{\partial\boldsymbol{x}}\cdot\boldsymbol{u}
\Bigg]
\\
&
\qquad \left(\boldsymbol{u}\cdot\boldsymbol{n}_j\right)
\,\mathrm{d}\boldsymbol{u}\,\mathrm{d}\boldsymbol{\xi}
\\
&+
\int_{\mathbb{R}^{D}} \int_{\boldsymbol{u}\cdot\boldsymbol{n}_j<0} \boldsymbol{\Psi}
\Bigg[
\left( \delta_{d,R}-\Delta t_Ge^{-\Delta t_G/\tau_R} \right)g_0^h
\\
&\quad+
\left( \delta_{e,R} + \frac{\Delta t_G^2}{2}e^{-\Delta t_G/\tau_R} \right)
\frac{\partial g^h}{\partial\boldsymbol{x}}\cdot\boldsymbol{u}
\Bigg]
\\
&
\qquad \left(\boldsymbol{u}\cdot\boldsymbol{n}_j\right)
\,\mathrm{d}\boldsymbol{u}\,\mathrm{d}\boldsymbol{\xi}.
\end{aligned}
\end{equation}
Here, $g_0^h$ is the interface Maxwellian corresponding to the hydrodynamic wave, and $\partial g^h/\partial\boldsymbol{x}$ is its reconstructed spatial derivative. Combining the equilibrium and analytic free transport wave fluxes with the net change obtained from the before-and-after particle tally gives the macroscopic update
\begin{equation}
\label{eq:ugkwp-macroscopic-update}
\boldsymbol W_i^{n+1} = \boldsymbol W_i^n -\frac{1}{\Omega_i} \sum_{j\in\mathcal M(i)}
\left( \boldsymbol F_j^{eq} + \boldsymbol F_j^{fr,wave}\right) s_j + \boldsymbol W_i^{fr,p}.
\end{equation}

\subsection{UGKWP algorithm}
\label{sec:ugkwp-algorithm}
\begin{figure*}[!t]
\centering
\includegraphics[width=0.82\textwidth]{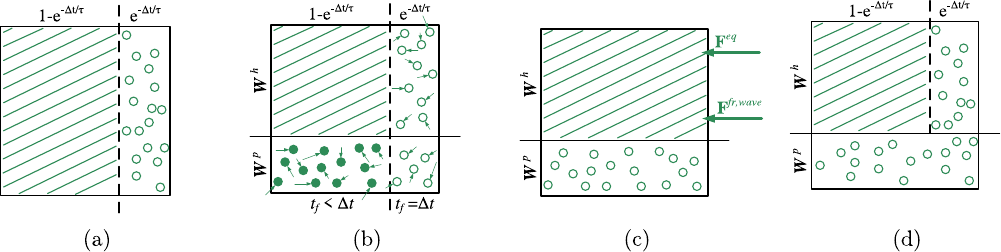}
\caption{\label{Fig:ugkwp_algorithm}Schematic illustration of the standard UGKWP update under GTS: (a) wave-particle decomposition and sampling, (b) wave and particle evolution, (c) particle tally and collision update, and (d) macroscopic update.}
\end{figure*}

The standard UGKWP update under GTS is summarized as follows.
\begin{enumerate}
\item \textbf{Initial state.} At the beginning of each time step, the cell-averaged conservative variables are decomposed according to Eq.~\eqref{eq:wave-particle-decomposition}. New collisionless particles are sampled from the hydrodynamic-wave state according to Eqs.~\eqref{eq:wave-resampling} and \eqref{eq:particle-sampling}. Together with the particles retained from the preceding step, they form the particle set at the current step as shown in Figs.~\ref{Fig:ugkwp_algorithm}(d). For the first step $n=0$, $\boldsymbol{W}^p = \boldsymbol{0}$ and $\boldsymbol{W}^h = \boldsymbol{W}^{n=0}$ as shown in Figs.~\ref{Fig:ugkwp_algorithm}(a).

\item \textbf{Free transport process.} For each retained particle, the free-transport time is determined by Eq.~\eqref{eq:mic1}, whereas each newly sampled particle is assigned \(t_{f,k}=\Delta t_G\). All particles are then transported according to Eq.~\eqref{eq:mic2}. The conservative quantities represented by particles before and after free transport are tallied according to Eq.~\eqref{eq:mic2b}, and their difference gives \(\boldsymbol{W}_i^{fr,p}\) in Eq.~\eqref{eq:mic3}. Particles with \(t_{f,k}<\Delta t_G\) are removed, whereas particles with \(t_{f,k}=\Delta t_G\) are retained. The analytic free-transport wave flux is evaluated using Eq.~\eqref{eq:free-wave-flux}.

\item \textbf{Collision process.} The equilibrium wave flux is evaluated using Eq.~\eqref{eq:feq}. The cell-averaged conservative variables are updated according to Eq.~\eqref{eq:ugkwp-macroscopic-update}, completing one UGKWP time step under GTS.

\item \textbf{Preparation for the next time step.} The updated cell-averaged conservative variables and the retained collisionless particles form the initial state for the next time step.
\end{enumerate}

In the standard GTS formulation, every cell uses the same observation scale $\Delta t_G$. This common scale affects the time-integration coefficients of the interface wave fluxes and controls the wave-particle decomposition through the local ratio $\Delta t_G/\tau_i$. For computations on nonuniform meshes or multiscale flow fields, however, the time step controlled by the CFL condition can vary substantially among cells because it depends on both the cell size and the local flow state. Replacing $\Delta t_G$ with the local time step $\Delta t_i$ therefore makes the observation scale spatially varying in both wave and particle evolution. Accordingly, the deterministic wave flux must remain consistent across each interface, and stochastic particles must satisfy the time-averaged interfacial flux balance across neighboring cells with different observation scales. In the present LTS construction, the equilibrium and analytic free transport wave fluxes on each side of an interface are integrated over the corresponding local time step and normalized by that time step to obtain the time-averaged interfacial wave fluxes. The sufficient condition for the time-averaged flux balance of stochastic particles is derived in the next section.

\section{Local time stepping and conservation analysis}
\label{sec:conservation-analysis}

For the explicit UGKWP discretization used in the steady flow simulation, the local time step of cell \(i\) is initialized as
\begin{equation}
\label{eq:cfl}
\Delta t_i = N_{\mathrm{CFL}} \frac{\Omega_i}{\sum\limits_{j\in\mathcal{M}(i)} \left[\left(|\boldsymbol{U}_i|+3\sqrt{RT_i}\right)s_j \right]},
\end{equation}
where $N_{\mathrm{CFL}} \in (0,1)$ is the CFL number. The quantities \(\boldsymbol U_i\) and \(T_i\) are evaluated from the initial flow field. The corresponding global time step is the minimum of the local time steps,
\begin{equation}
\label{eq:gts}
\Delta t_G=\min_i(\Delta t_i).
\end{equation}
The initialized local time steps remain fixed throughout the steady calculation. Thus, all cells use \(\Delta t_G\) in GTS, whereas cell \(i\) uses \(\Delta t_i\) in LTS.

\subsection{Time-averaged interfacial particle-flux balance}
\label{sec:time-averaged-particle-flux-balance}
Under GTS, a particle crossing an interface carries the same mass on both sides and is recorded in the flux evaluations of both cells using the same \(\Delta t_G\). The corresponding stochastic-particle flux is therefore balanced across the interface. Under LTS, adjacent cells may use different time steps. We derive a sufficient condition for balancing the time-averaged stochastic-particle flux across their shared interface. Consider two adjacent cells \(L\) and \(R\) sharing interface \(j\). For analysis, let \(\boldsymbol n_j\) point from \(L\) to \(R\). We consider particles crossing from \(L\) to \(R\) and treat those crossing in the opposite direction analogously. Over \(M\) time steps, the corresponding time-averaged stochastic-particle fluxes recorded on the two sides are
\begin{equation}
\label{eq:avg_flux_1}
\begin{aligned}
\overline{\boldsymbol{F}}_L^{fr,p}
&=
\frac{
\displaystyle
\sum_{\ell=1}^{M}
\left(
\sum_{k\in K_{\ell}}
m_{L,k}^{(\ell)}\boldsymbol{\Psi}_k
\right)
}{
\displaystyle
s_j\sum_{\ell=1}^{M}\Delta t_{L,\ell}
},\\
\overline{\boldsymbol{F}}_R^{fr,p}
&=
\frac{
\displaystyle
\sum_{\ell=1}^{M}
\left(
\sum_{k\in K_{\ell}}
m_{R,k}^{(\ell)}\boldsymbol{\Psi}_k
\right)
}{
\displaystyle
s_j\sum_{\ell=1}^{M}\Delta t_{R,\ell}
}.
\end{aligned}
\end{equation}

Here $K_{\ell}$ is the set of crossing particles during step $\ell$, while $m_{L,k}^{(\ell)}$ and $m_{R,k}^{(\ell)}$ are the masses of particle $k$ recorded on the $L$ and $R$ sides of the interface during step $\ell$, respectively. The molecular velocity \(\boldsymbol{u}_k\) and equivalent velocity \(\boldsymbol{\xi}_k\) remain unchanged during the interface transfer. The required time-averaged particle-flux balance is
\begin{equation}
\label{eq:avg_flux_equal_1}
\overline{\boldsymbol{F}}_L^{fr,p}=\overline{\boldsymbol{F}}_R^{fr,p}.
\end{equation}
To formulate the balance condition, let $\mathcal{V}=\mathbb{R}^{3}\times\mathbb{R}^{D}$ denote the $(\boldsymbol{u},\boldsymbol{\xi})$ space and partition $\mathcal{V}$ into bounded regions $\varpi$. Define the partition scale as
\[
h_{\mathcal V} = \sup_{\varpi} \sup_{\substack{(\boldsymbol{u},\boldsymbol{\xi})\in\varpi\\
(\boldsymbol{u}',\boldsymbol{\xi}')\in\varpi}}
\left\| (\boldsymbol{u},\boldsymbol{\xi}) - (\boldsymbol{u}',\boldsymbol{\xi}') \right\|.
\]
For each region $\varpi$, choose a representative point $(\boldsymbol{u}_{\varpi},\boldsymbol{\xi}_{\varpi})$. The crossing particles in region $\varpi$ during step $\ell$ are
\[
K_{\ell,\varpi} = \left\{k\in K_\ell: (\boldsymbol{u}_k,\boldsymbol{\xi}_k)\in\varpi \right\}.
\]
For every nonempty $K_{\ell,\varpi}$,
\[
\max_{k\in K_{\ell,\varpi}} \left\| \boldsymbol{\Psi}_k-\boldsymbol{\Psi}_{\varpi} \right\| \rightarrow 0
\qquad
\text{as }h_{\mathcal V}\rightarrow0.
\]
For these crossing particles, the total masses recorded on the \(L\) side before crossing and on the \(R\) side after crossing, respectively, are
\begin{equation}
\label{eq:mass_dt_balance_0}
\begin{aligned}
\widetilde{m}_{L,\ell,\varpi}
&=
\sum_{k\in K_{\ell,\varpi}}m_{L,k}^{(\ell)},\\
\widetilde{m}_{R,\ell,\varpi}
&=
\sum_{k\in K_{\ell,\varpi}}m_{R,k}^{(\ell)}.
\end{aligned}
\end{equation}
The time-averaged stochastic-particle fluxes associated with these crossing particles are
\begin{equation}
\label{eq:avg_flux_2}
\begin{aligned}
\overline{\boldsymbol{F}}_{L,\varpi}^{fr,p} &= \frac{\displaystyle \boldsymbol{\Psi}_{\varpi}\sum_{\ell=1}^{M}\widetilde{m}_{L,\ell,\varpi}}
{
\displaystyle s_j\sum_{\ell=1}^{M}\Delta t_{L,\ell}},\\
\overline{\boldsymbol{F}}_{R,\varpi}^{fr,p} &= \frac{\displaystyle \boldsymbol{\Psi}_{\varpi}\sum_{\ell=1}^{M}\widetilde{m}_{R,\ell,\varpi}}
{\displaystyle s_j\sum_{\ell=1}^{M}\Delta t_{R,\ell}}.
\end{aligned}
\end{equation}
The complete time-averaged stochastic-particle fluxes are recovered by summing over the velocity-space partition and taking the refinement limit,
\[
\begin{aligned}
  \overline{\boldsymbol{F}}_{L}^{fr,p} &= \lim_{h_{\mathcal V}\rightarrow0} \sum_{\varpi} \overline{\boldsymbol{F}}_{L,\varpi}^{fr,p},\\
  \overline{\boldsymbol{F}}_{R}^{fr,p} &= \lim_{h_{\mathcal V}\rightarrow0} \sum_{\varpi} \overline{\boldsymbol{F}}_{R,\varpi}^{fr,p}.
\end{aligned}
\]

A sufficient construction for Eq.~\eqref{eq:avg_flux_equal_1} is to impose the flux-balance condition in every velocity-space region,
\begin{equation}
\label{eq:avg_flux_equal_2}
\overline{\boldsymbol{F}}_{L,\varpi}^{fr,p}
=
\overline{\boldsymbol{F}}_{R,\varpi}^{fr,p}.
\end{equation}
Since \(s_j\) and \(\boldsymbol{\Psi}_{\varpi}\) are the same in the two regional flux evaluations, Eq.~\eqref{eq:avg_flux_equal_2} is satisfied when
\begin{equation}
\label{eq:mass_dt_balance}
\frac{
\displaystyle\sum_{\ell=1}^{M}
\widetilde{m}_{L,\ell,\varpi}
}{
\displaystyle\sum_{\ell=1}^{M}\Delta t_{L,\ell}
}
=
\frac{
\displaystyle\sum_{\ell=1}^{M}
\widetilde{m}_{R,\ell,\varpi}
}{
\displaystyle\sum_{\ell=1}^{M}\Delta t_{R,\ell}
}.
\end{equation}
Under this sufficient construction, the particle mass transported per unit accumulated time is balanced within every velocity-space region. For a finite partition, the result is subject to the corresponding velocity-space partition error. Since \(\boldsymbol{u}_k\) and \(\boldsymbol{\xi}_k\) are recorded explicitly for each particle, this error vanishes as \(h_{\mathcal V} \to 0\).

Suppose that the two cells use fixed positive local time steps, such that \(\Delta t_{L,\ell}=\Delta t_L\) and \(\Delta t_{R,\ell}=\Delta t_R\) for every \(\ell\). For region \(\varpi\), a sufficient particle-level scaling applied to each crossing particle is
\begin{equation}
\label{eq:mass_dt_balance_1}
m_{R,k}^{(\ell)} = m_{L,k}^{(\ell)} \frac{\Delta t_R}{\Delta t_L},   \qquad   k\in K_{\ell,\varpi}.
\end{equation}
Summing this relation over the same set of crossing particles gives
\begin{equation}
\label{eq:mass_dt_balance_2}
\widetilde m_{R,\ell,\varpi} = \widetilde m_{L,\ell,\varpi} \frac{\Delta t_R}{\Delta t_L}.
\end{equation}
Since the accumulated local times of cells \(L\) and \(R\) are \(M\Delta t_L\) and \(M\Delta t_R\), respectively, substitution into Eq.~\eqref{eq:avg_flux_2} gives
\begin{equation}
\label{eq:fixed-step-flux-balance}
\begin{aligned}
\overline{\boldsymbol{F}}_{L,\varpi}^{fr,p} &= \frac{ \displaystyle \boldsymbol{\Psi}_{\varpi} \sum_{\ell=1}^{M}\widetilde{m}_{L,\ell,\varpi} }{ s_jM\Delta t_L }\\
        &= \frac{ \displaystyle \boldsymbol{\Psi}_{\varpi} \sum_{\ell=1}^{M}\widetilde{m}_{R,\ell,\varpi} }{ s_jM\Delta t_R }
        = \overline{\boldsymbol{F}}_{R,\varpi}^{fr,p}.
\end{aligned}
\end{equation}
Consequently, the proportional scaling balances all components of the particle flux within each velocity-space region. Summing over the partition and taking \(h_{\mathcal V}\rightarrow0\) gives the complete time-averaged stochastic-particle flux balance for mass, momentum, and energy. Hence, the particle-mass scaling in Eq.~\eqref{eq:mass_dt_balance_1}, together with fixed positive local time steps, provides a sufficient condition for interfacial time-averaged particle-flux balance. The flux-balance argument imposes no additional restriction on the ratio \(\Delta t_R/\Delta t_L\) or on the spatial variation of the fixed local time steps. This statement concerns interfacial time-averaged particle-flux balance and does not replace the CFL or stability requirements used to determine the individual local time steps. Applying the same construction in the opposite direction yields time-averaged stochastic-particle flux balance across the interface.

To examine why stepwise proportional mass scaling alone is insufficient when the local time steps vary during the computation, consider the stepwise regional transported-mass scaling,
\begin{equation}
\label{eq:time-varying-mass-scaling}
\widetilde{m}_{R,\ell,\varpi} = \widetilde{m}_{L,\ell,\varpi} \frac{\Delta t_{R,\ell}}{\Delta t_{L,\ell}}.
\end{equation}
For a velocity-space region satisfying \(\sum_{\ell=1}^{M}\widetilde{m}_{L,\ell,\varpi}>0\), substituting Eq.~\eqref{eq:time-varying-mass-scaling} into Eq.~\eqref{eq:mass_dt_balance} yields the additional condition
\begin{equation}
\label{eq:time-varying-balance-condition}
\frac{ \displaystyle\sum_{\ell=1}^{M}\Delta t_{R,\ell} }{ \displaystyle\sum_{\ell=1}^{M}\Delta t_{L,\ell} }
=
\frac{ \displaystyle\sum_{\ell=1}^{M} \widetilde{m}_{L,\ell,\varpi} \frac{\Delta t_{R,\ell}}{\Delta t_{L,\ell}} }{ \displaystyle\sum_{\ell=1}^{M} \widetilde{m}_{L,\ell,\varpi} }.
\end{equation}
To expose the mismatch explicitly, assume that \(\widetilde{m}_{L,\ell,\varpi}\) remains constant over the \(M\) steps. Then Eq.~\eqref{eq:time-varying-balance-condition} reduces to
\begin{equation}
\label{eq:time-varying-unweighted-condition}
\frac{ \displaystyle\sum_{\ell=1}^{M}\Delta t_{R,\ell} }{ \displaystyle\sum_{\ell=1}^{M}\Delta t_{L,\ell} }
=
\frac{1}{M} \sum_{\ell=1}^{M} \frac{\Delta t_{R,\ell}}{\Delta t_{L,\ell}}.
\end{equation}
A general time-step sequence does not necessarily satisfy the balance condition in Eq.~\eqref{eq:time-varying-unweighted-condition}. For example, with \(M=2\), \((\Delta t_{L,1},\Delta t_{L,2})=(1,2)\) and \((\Delta t_{R,1},\Delta t_{R,2})=(2,2)\), its left- and right-hand sides are \(4/3\) and \(3/2\), respectively. Therefore, stepwise proportional mass scaling alone does not guarantee the multi-step time-averaged particle-flux balance when the local time steps vary during the computation. An adaptive DSMC procedure used empirical controls, including limiting the time-step update and smoothing the local time steps~\cite{boyd2015jcp}, thereby suppressing abrupt variations of the interface time-step ratio. Such controls, however, do not establish the multi-step balance condition in Eq.~\eqref{eq:time-varying-balance-condition}.

The preceding analysis provides a sufficient construction for time-averaged particle-flux balance when particles cross between cells with unequal local observation scales. This sufficient construction is applied to the UGKWP-LTS implementation in the following subsection.

\subsection{UGKWP-LTS implementation}
\label{sec:implementation}
The UGKWP-LTS update uses the equilibrium wave flux and the analytic free transport wave flux together with the before-and-after tally of stochastic particles. The hydrodynamic wave in cell $i$ is defined by $\boldsymbol{W}_i^h=\boldsymbol{W}_i-\boldsymbol{W}_i^p$. The wave-particle decomposition uses the local ratio \(\Delta t_i/\tau_i\), so that the collisionless part is evaluated as \(e^{-\Delta t_i/\tau_i}\), rather than \(e^{-\Delta t_G/\tau_i}\). Under LTS, the conservative variables represented by the newly sampled collisionless particles are \(\boldsymbol{W}_i^{hp}=\boldsymbol{W}_i^h e^{-\Delta t_i/\tau_i}\), and each such particle is assigned \(t_{f,k}=\Delta t_i\). Let $\delta_\alpha(\Delta t_i,\tau_i)$, $\alpha\in\{a,b,c,d,e\}$, denote the corresponding time-integration coefficient in Eq.~\eqref{eq:time-integration-coefficients} evaluated with local time step $\Delta t_i$ and local relaxation time $\tau_i$. For each side $\sigma\in\{L,R\}$ of an interface, the LTS coefficients are
\begin{equation}
\label{eq:lts-side-dependent-coefficients}
\delta_{\alpha,\sigma}
=
\delta_\alpha(\Delta t_\sigma,\tau_\sigma),
\qquad
\alpha\in\{a,b,c,d,e\}.
\end{equation}
Each coefficient is normalized by the corresponding cell-side time step,
\begin{equation}
\label{eq:lts-normalized-coefficients}
\overline{\delta}_{\alpha,\sigma}
=
\frac{\delta_{\alpha,\sigma}}{\Delta t_\sigma},
\qquad
\alpha\in\{a,b,c,d,e\}.
\end{equation}
At a physical boundary adjacent to fluid cell \(i\), both half-space wave fluxes use \(\Delta t_i\). The left and right wave fluxes are integrated over their respective local time steps and then normalized by those time steps to obtain time-averaged fluxes. The corresponding time-averaged interfacial equilibrium wave flux is
\begin{equation}
\label{eq:lts-feq}
\begin{aligned}
\overline{\boldsymbol{F}}^{eq}_{j}={}&
\int_{\mathbb{R}^{D}}
\int_{\boldsymbol{u}\cdot\boldsymbol{n}_{j}>0}
\boldsymbol{\Psi}
\left(
\overline{\delta}_{a,L}g_0
+\overline{\delta}_{b,L}
\frac{\partial g}{\partial\boldsymbol{x}}\cdot\boldsymbol{u}
+\overline{\delta}_{c,L}
\frac{\partial g}{\partial t}
\right)
\\
&\qquad
\left(\boldsymbol{u}\cdot\boldsymbol{n}_{j}\right)
\,\mathrm{d}\boldsymbol{u}\,\mathrm{d}\boldsymbol{\xi}
\\
&+
\int_{\mathbb{R}^{D}}
\int_{\boldsymbol{u}\cdot\boldsymbol{n}_{j}<0}
\boldsymbol{\Psi}
\left(
\overline{\delta}_{a,R}g_0
+\overline{\delta}_{b,R}
\frac{\partial g}{\partial\boldsymbol{x}}\cdot\boldsymbol{u}
+\overline{\delta}_{c,R}
\frac{\partial g}{\partial t}
\right)
\\
&\qquad
\left(\boldsymbol{u}\cdot\boldsymbol{n}_{j}\right)
\,\mathrm{d}\boldsymbol{u}\,\mathrm{d}\boldsymbol{\xi}.
\end{aligned}
\end{equation}

Replacing $\Delta t_G$ in Eq.~\eqref{eq:free-wave-flux} with the corresponding cell-side time step $\Delta t_\sigma$
and applying the same normalization gives the time-averaged interfacial analytic free-transport wave flux
\begin{equation}
\label{eq:ffrwave}
\begin{aligned}
\overline{\boldsymbol{F}}^{fr,wave}_{j}={}&
\int_{\mathbb{R}^{D}} \int_{\boldsymbol{u}\cdot\boldsymbol{n}_{j}>0} \boldsymbol{\Psi}
\Bigg[
\left(\overline{\delta}_{d,L} - e^{-\Delta t_L/\tau_L} \right) g_0^h
\\
&\quad+
\left(\overline{\delta}_{e,L} + \frac{\Delta t_L}{2}e^{-\Delta t_L/\tau_L}\right)
\frac{\partial g^h}{\partial\boldsymbol{x}}\cdot\boldsymbol{u}
\Bigg]
\\
&\qquad
\left(\boldsymbol{u}\cdot\boldsymbol{n}_{j}\right)
\,\mathrm{d}\boldsymbol{u}\,\mathrm{d}\boldsymbol{\xi}
\\
&+
\int_{\mathbb{R}^{D}} \int_{\boldsymbol{u}\cdot\boldsymbol{n}_{j}<0} \boldsymbol{\Psi}
\Bigg[
\left( \overline{\delta}_{d,R} - e^{-\Delta t_R/\tau_R} \right) g_0^h
\\
&\quad+
\left( \overline{\delta}_{e,R} + \frac{\Delta t_R}{2}e^{-\Delta t_R/\tau_R} \right)
\frac{\partial g^h}{\partial\boldsymbol{x}}\cdot\boldsymbol{u}
\Bigg]
\\
&\qquad
\left(\boldsymbol{u}\cdot\boldsymbol{n}_{j}\right)
\,\mathrm{d}\boldsymbol{u}\,\mathrm{d}\boldsymbol{\xi}.
\end{aligned}
\end{equation}

For a particle retained from the preceding step and initially located in cell $i$, the free-transport time is determined using the local observation scale,
\begin{equation}
\label{eq:lts-free-transport-time}
t_{f,k}=\min\left[-\tau_i\ln(\epsilon_k),\Delta t_i\right].
\end{equation}
 Newly sampled collisionless particles are assigned \(t_{f,k}=\Delta t_i\). When a particle crosses from cell $L$ to cell $R$, its mass is rescaled according to Eq.~\eqref{eq:mass_dt_balance_1}. The remaining free-transport time is also rescaled by the same ratio. If $t_{r,k,L}$ and $t_{r,k,R}$ denote the remaining free-transport times immediately before and after the transfer, respectively, the corresponding rule is
\begin{equation}
\label{eq:proportional_scaling_tf}
t_{r,k,R}=t_{r,k,L}\frac{\Delta t_R}{\Delta t_L}.
\end{equation}

Using the time-averaged wave fluxes and the before-and-after particle tally $\boldsymbol{W}_i^{fr,p}$ defined in Eq.~\eqref{eq:mic3}, the cell-averaged conservative variables are updated as
\begin{equation}
\label{eq:renew}
\begin{aligned}
\boldsymbol{W}^{n+1}_i = \boldsymbol{W}^{n}_i - \frac{\Delta t_i}{\Omega_i} \sum_{j\in\mathcal{M}(i)}
\left( \overline{\boldsymbol F}^{eq}_j + \overline{\boldsymbol F}^{fr,wave}_j \right) s_j + \boldsymbol W_i^{fr,p}.
\end{aligned}
\end{equation}

When $\Delta t_L=\Delta t_R=\Delta t_G$, the normalized coefficients satisfy $\Delta t_G\overline{\delta}_{\alpha,\sigma}=\delta_{\alpha,\sigma}$. Consequently, $\Delta t_G\overline{\boldsymbol{F}}^{eq}_j=\boldsymbol{F}^{eq}_j$ and $\Delta t_G\overline{\boldsymbol{F}}^{fr,wave}_j=\boldsymbol{F}^{fr,wave}_j$, while the particle-mass scaling reduces to $m_{R,k}^{(\ell)}=m_{L,k}^{(\ell)}$ for every step $\ell$. The same before-and-after particle tally $\boldsymbol{W}_i^{fr,p}$ as in GTS is then used. Equation~\eqref{eq:renew} therefore recovers the GTS update in Eq.~\eqref{eq:ugkwp-macroscopic-update}.

\section{Numerical Verification}
\label{sec:verification}
The UGKWP-LTS method is evaluated using a hypersonic cylinder flow and a hypersonic flat-plate flow. All GTS and LTS calculations are initialized with the corresponding uniform freestream state. Published UGKS solutions are used to evaluate the cylinder-flow quantities, while paired GTS calculations serve as the references for convergence and wall-clock performance. Step-count and wall-clock speedups are reported separately. Unless otherwise stated, the wall-clock measurements were obtained using 28 cores of an Intel Xeon Gold 6348 processor.

\subsection{Hypersonic Flow Past a Cylinder}
\label{sec:cylinder}
The hypersonic flow past a circular cylinder is considered at $\mathrm{Ma}=5$. The cylinder radius $R=1$ is used as the reference length, and the wall and freestream temperatures are set to $T_w=T_\infty=1$. The freestream Knudsen number is defined as $\mathrm{Kn}=\lambda_\infty/R$, and the cases at $\mathrm{Kn}=0.01$ and $0.1$ are considered. The working gas is argon, modeled by the Shakhov model with $\mathrm{Pr}=1.0$ and the variable hard sphere (VHS) model with $\omega=0.81$. The diffuse-reflection boundary condition is imposed on the cylinder surface in both cases. Both GTS and LTS calculations use a CFL number of $0.8$ and $N_{\mathrm{ref}}=100$. For both time-stepping strategies, the $\mathrm{Kn}=0.01$ case uses an unstructured mesh with $16{,}100$ cells and a near-wall cell height of $0.0005R$, whereas the $\mathrm{Kn}=0.1$ case uses an unstructured mesh with $7{,}420$ cells and a near-wall cell height of $0.01R$.

\begin{figure}[tbp]
\centering
\subfigure[$\mathrm{Kn}=0.01$, $C_p$]{\includegraphics[width=0.23\textwidth]{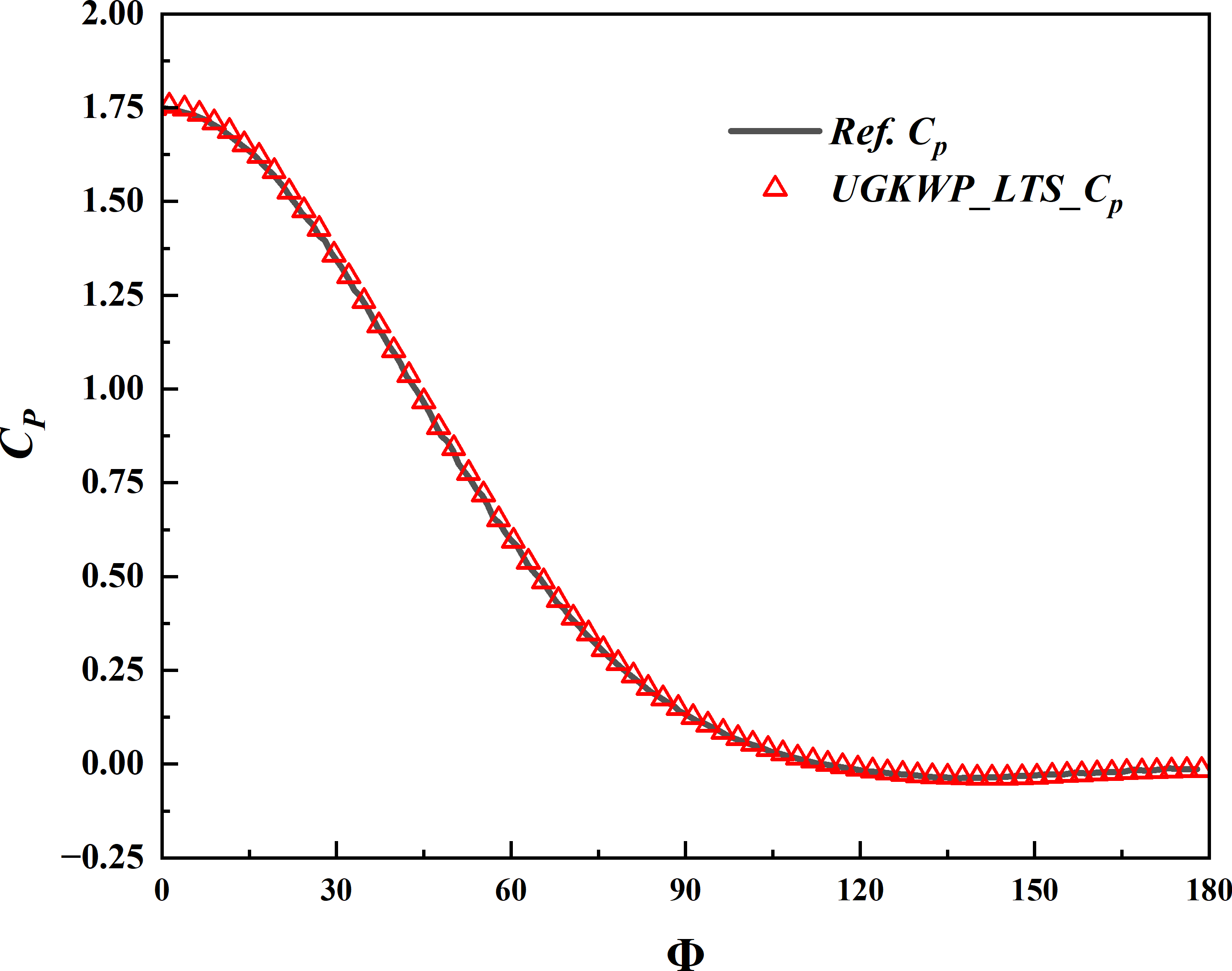}}
\subfigure[$\mathrm{Kn}=0.01$, $C_F$ and $C_Q$]{\includegraphics[width=0.23\textwidth]{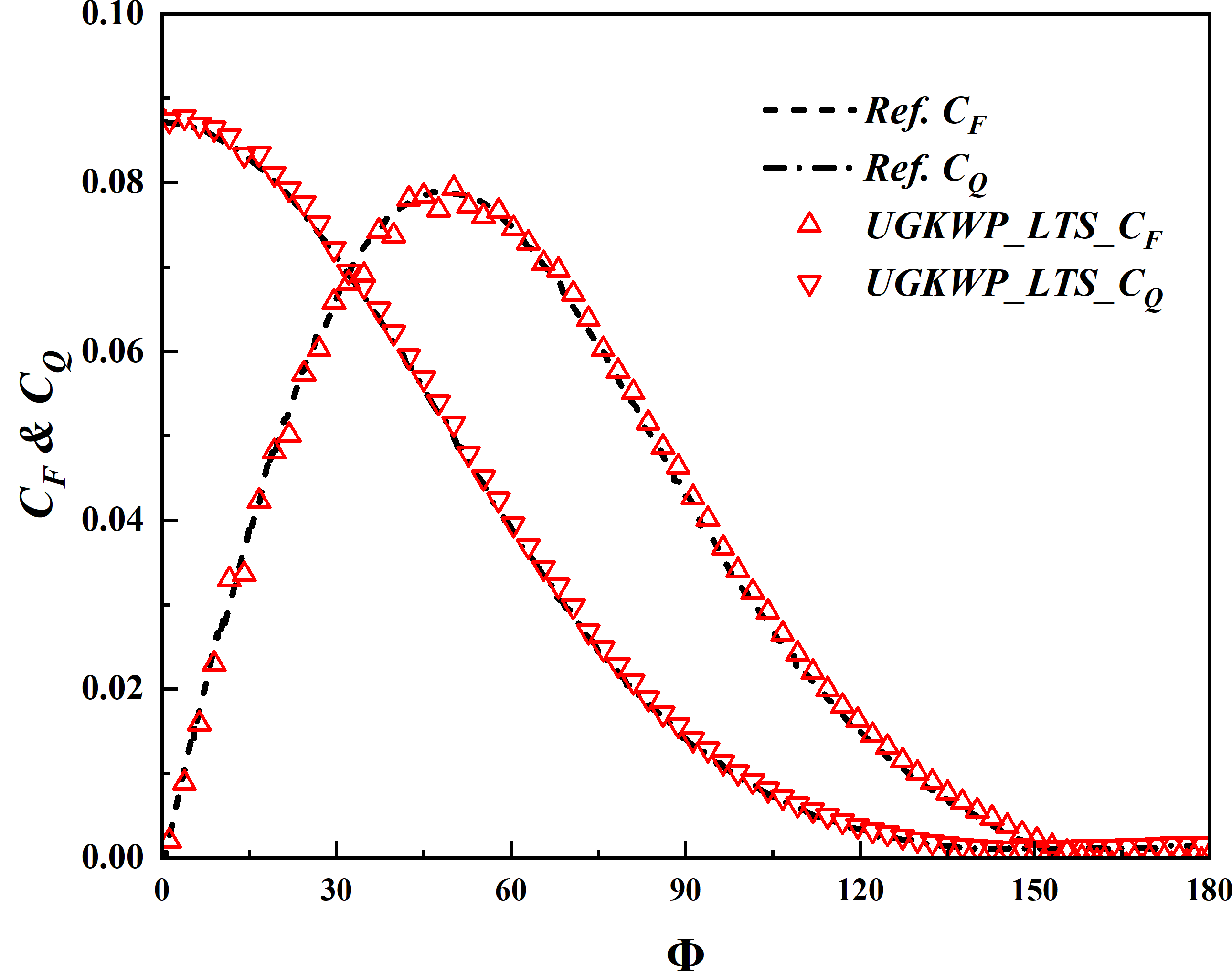}}\\
\subfigure[$\mathrm{Kn}=0.1$, $C_p$]{\includegraphics[width=0.23\textwidth]{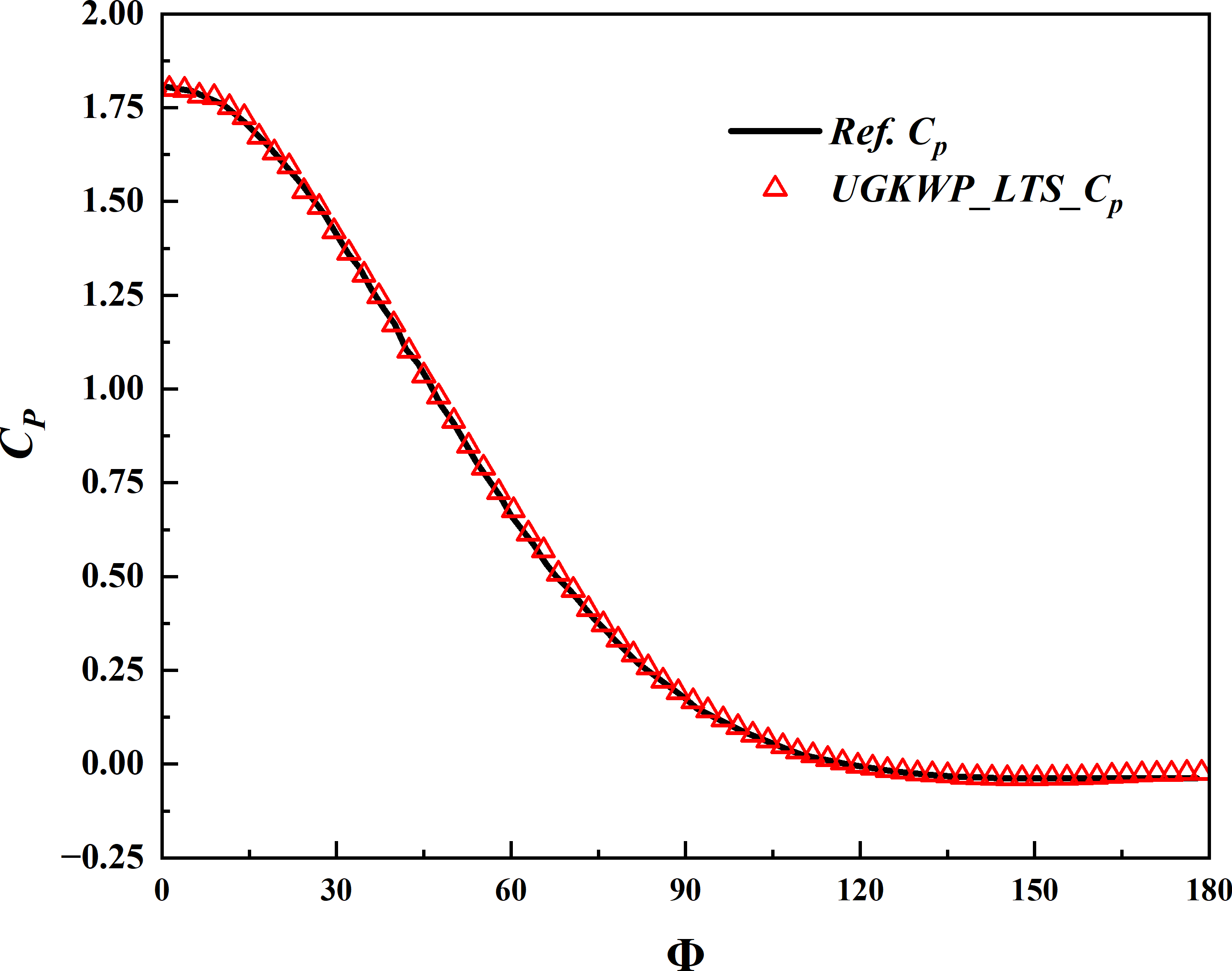}}
\subfigure[$\mathrm{Kn}=0.1$, $C_F$ and $C_Q$]{\includegraphics[width=0.23\textwidth]{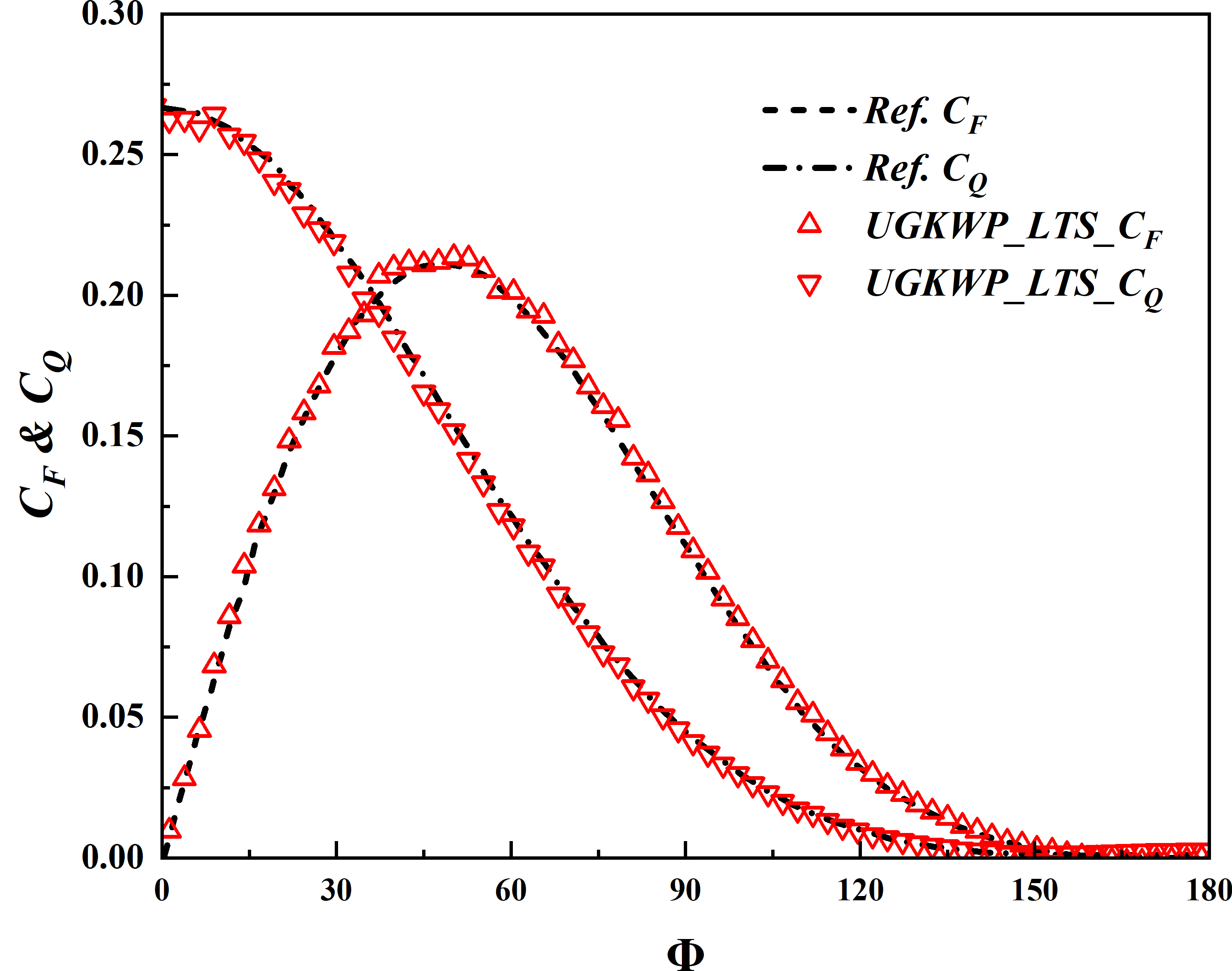}}
\caption{\label{Fig:cyl_wall}
Wall pressure, shear stress, and heat flux coefficients at $\mathrm{Kn}=0.01$ and $0.1$.}

\centering
\subfigure[$\mathrm{Kn}=0.01$, density]{\includegraphics[width=0.23\textwidth]{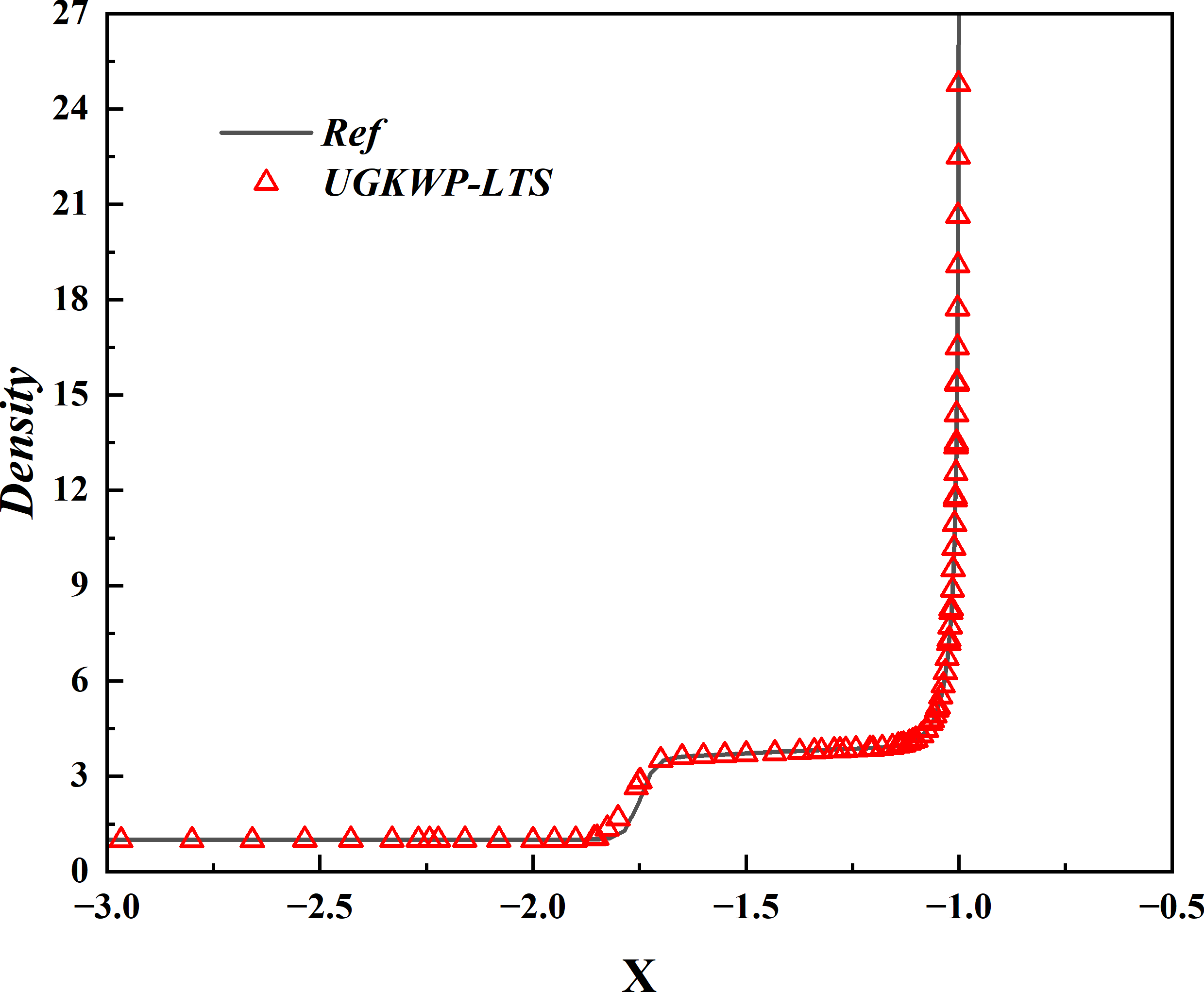}}
\subfigure[$\mathrm{Kn}=0.01$, velocity and temperature]{\includegraphics[width=0.23\textwidth]{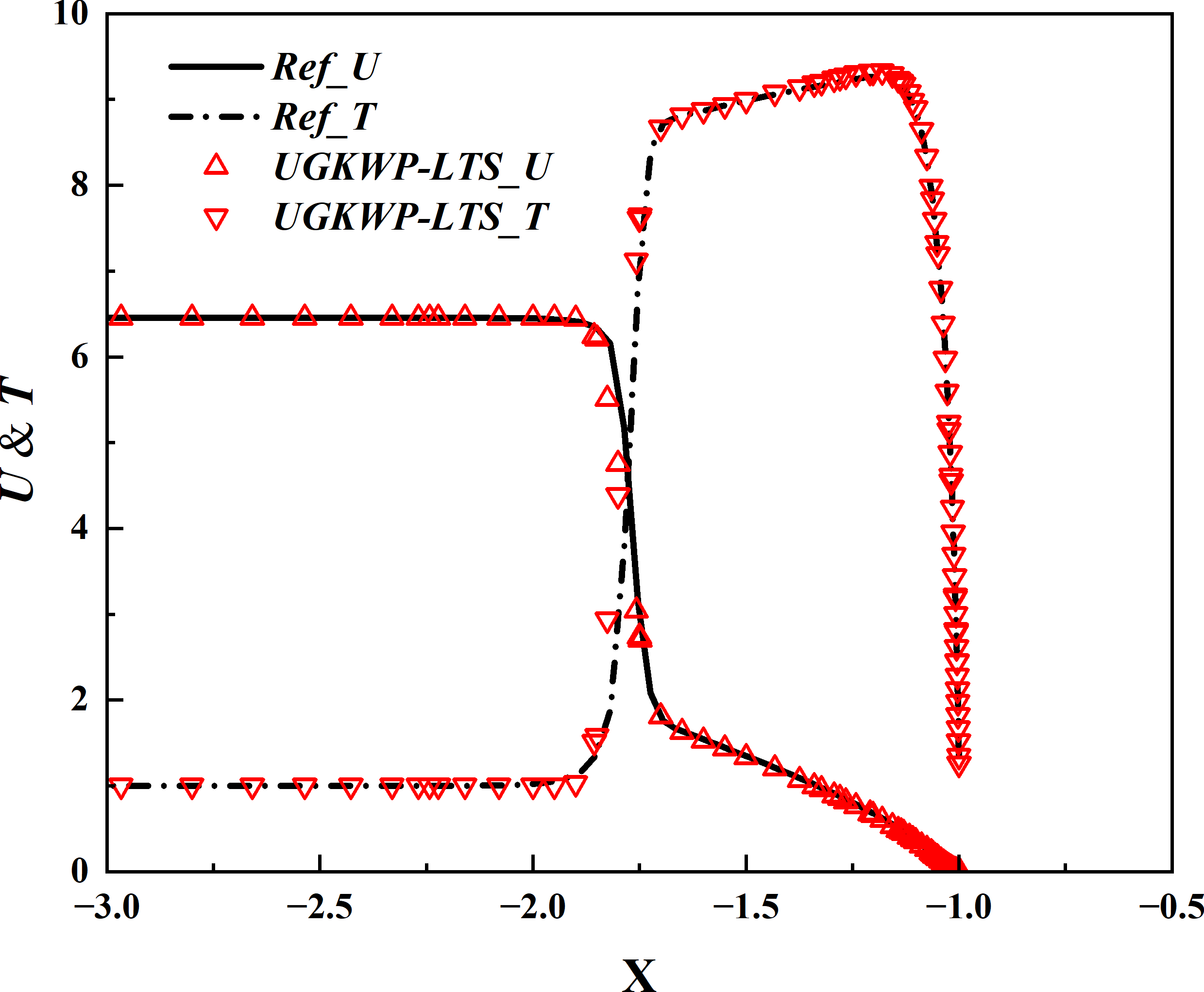}}\\
\subfigure[$\mathrm{Kn}=0.1$, density]{\includegraphics[width=0.23\textwidth]{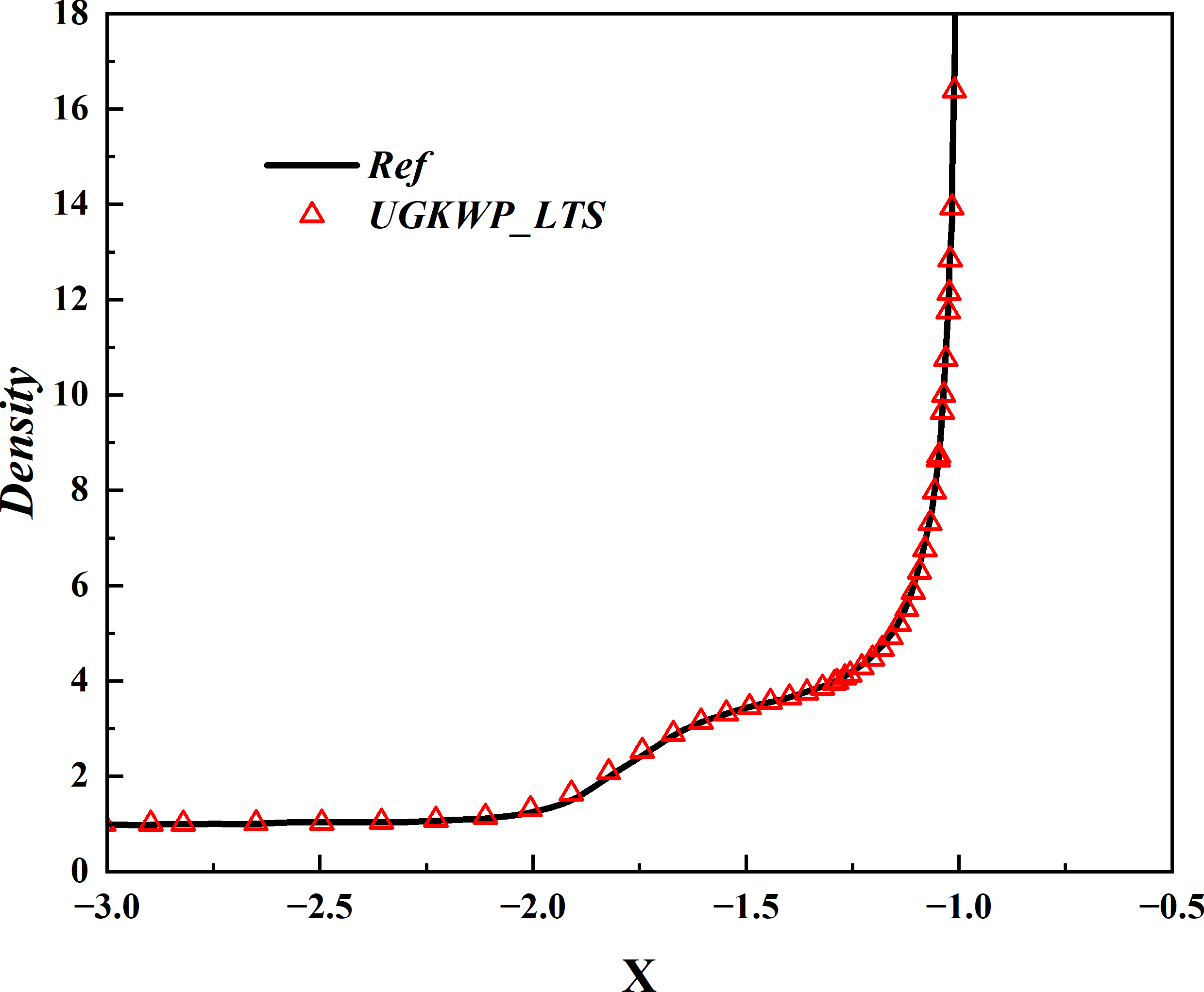}}
\subfigure[$\mathrm{Kn}=0.1$, velocity and temperature]{\includegraphics[width=0.23\textwidth]{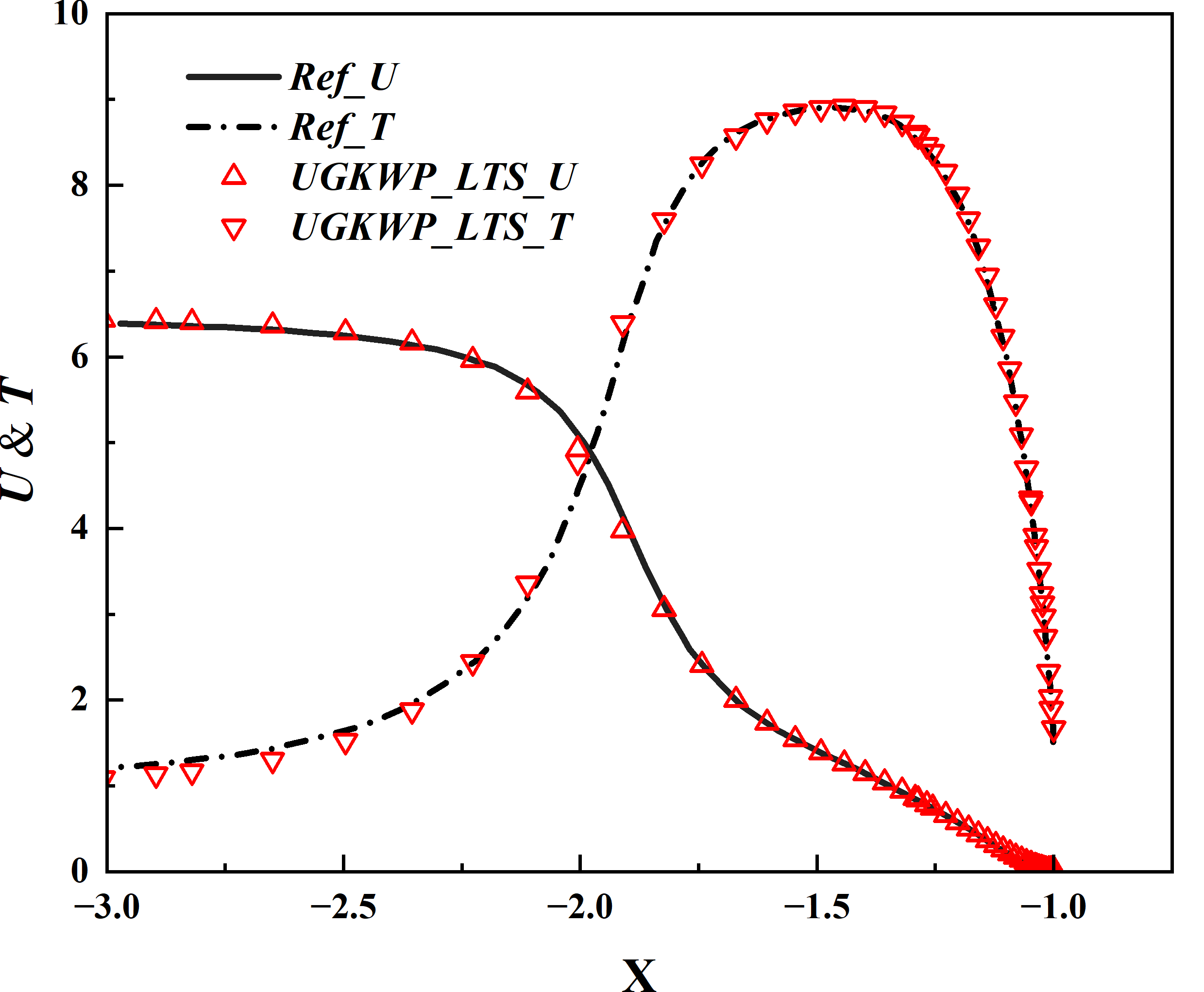}}
\caption{\label{Fig:cyl_stag}
Density, velocity, and temperature profiles along the stagnation line at $\mathrm{Kn}=0.01$ and $0.1$.}
\end{figure}

The wall quantities and stagnation-line profiles calculated by UGKWP-LTS are compared with the UGKS reference solutions~\cite{awp2}. The wall pressure, shear stress, and heat flux coefficients are defined as
\begin{equation}
\begin{aligned}
  C_p &= \frac{p_w-p_\infty}{\frac{1}{2}\rho_\infty |\boldsymbol{U}_\infty|^2},\\
  C_F &= \frac{\tau_w}{\frac{1}{2}\rho_\infty |\boldsymbol{U}_\infty|^2},\\
  C_Q &= \frac{q_w}{\frac{1}{2}\rho_\infty |\boldsymbol{U}_\infty|^3},
\end{aligned}
\end{equation}
where $p_w$, $\tau_w$, and $q_w$ denote the wall pressure, shear stress, and heat flux, respectively. Here, $p_\infty$, $\rho_\infty$, and $\boldsymbol{U}_\infty$ denote the freestream pressure, density, and velocity, respectively. Figures~\ref{Fig:cyl_wall} and~\ref{Fig:cyl_stag} compare the wall coefficients and stagnation-line profiles with the UGKS reference solutions, respectively. At both $\mathrm{Kn}=0.01$ and $0.1$, the UGKWP-LTS results agree closely with the reference solutions.

\begin{figure*}[t]
\centering
\subfigure[GTS, $\log_{10}(\Delta t_G/\tau_i)$]{
\includegraphics[width=0.23\textwidth,trim=3bp 3bp 3bp 3bp,clip]{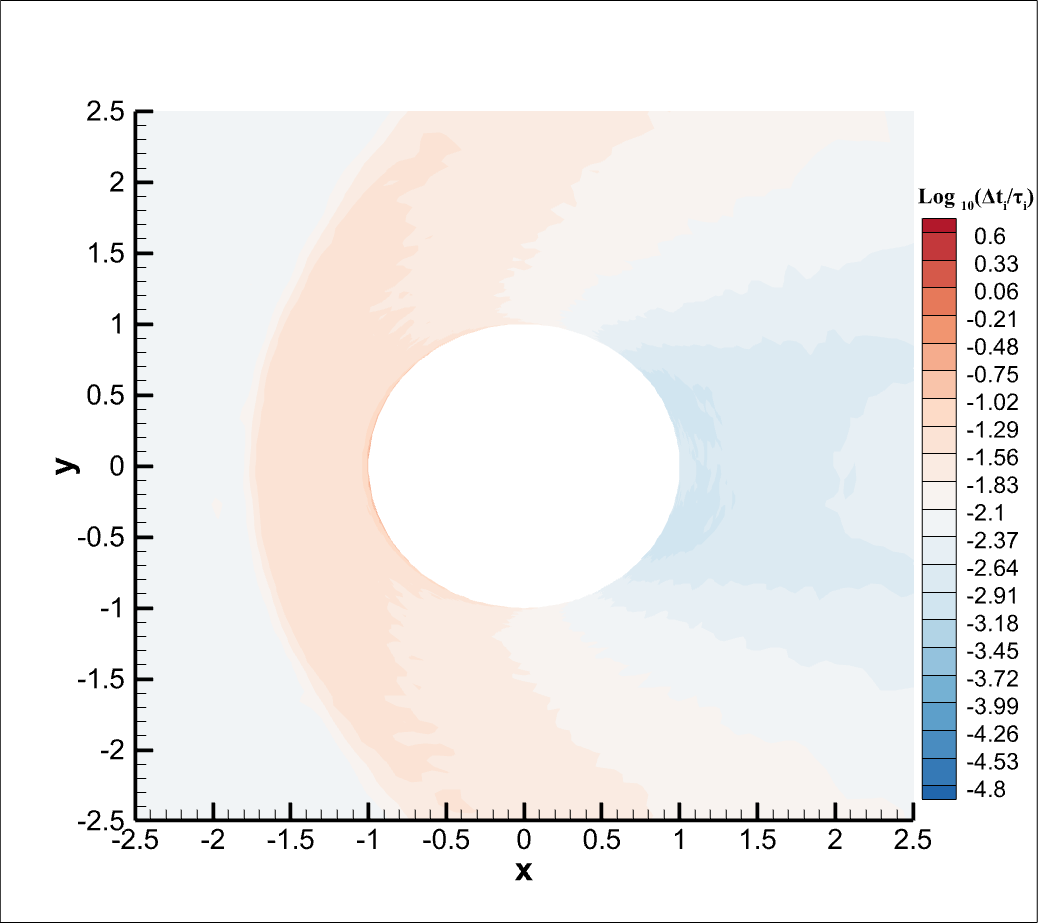}
}\hfill
\subfigure[LTS, $\log_{10}(\Delta t_i/\tau_i)$]{
\includegraphics[width=0.23\textwidth,trim=3bp 3bp 3bp 3bp,clip]{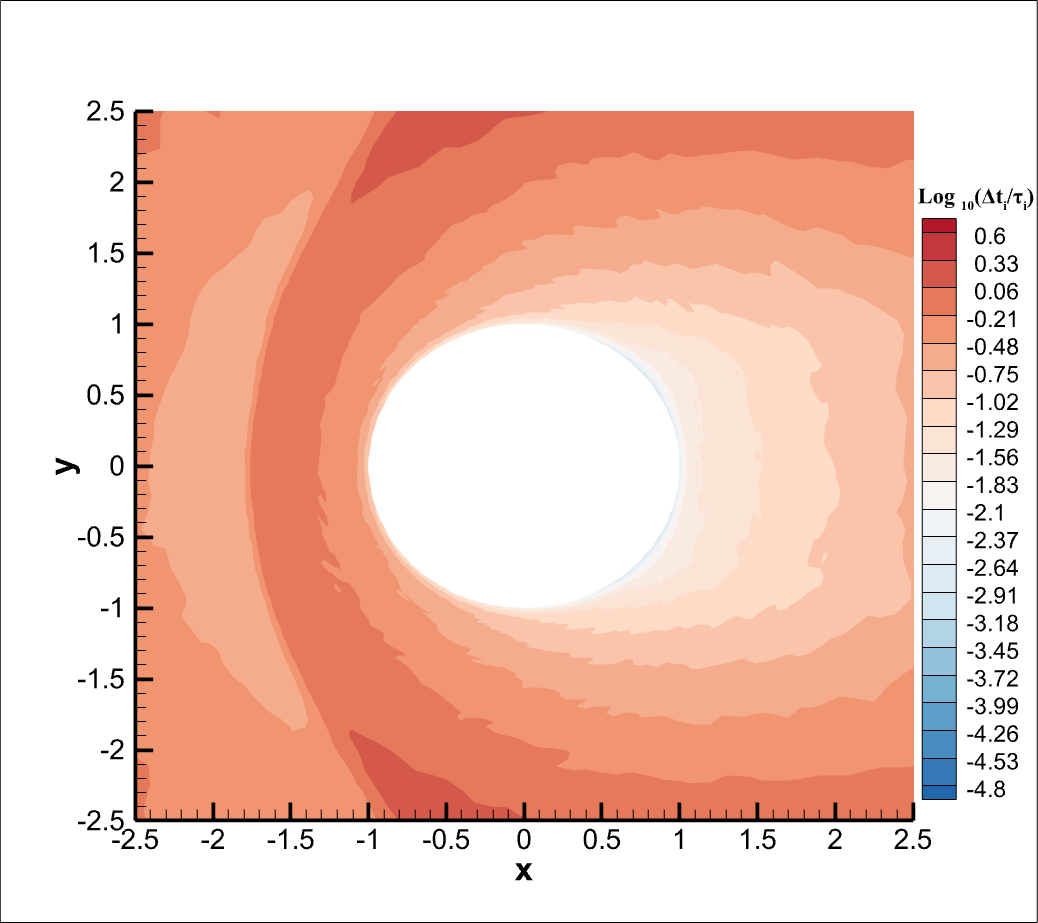}
}\hfill
\subfigure[GTS, particle mass ratio]{
\includegraphics[width=0.23\textwidth,trim=3bp 3bp 3bp 3bp,clip]{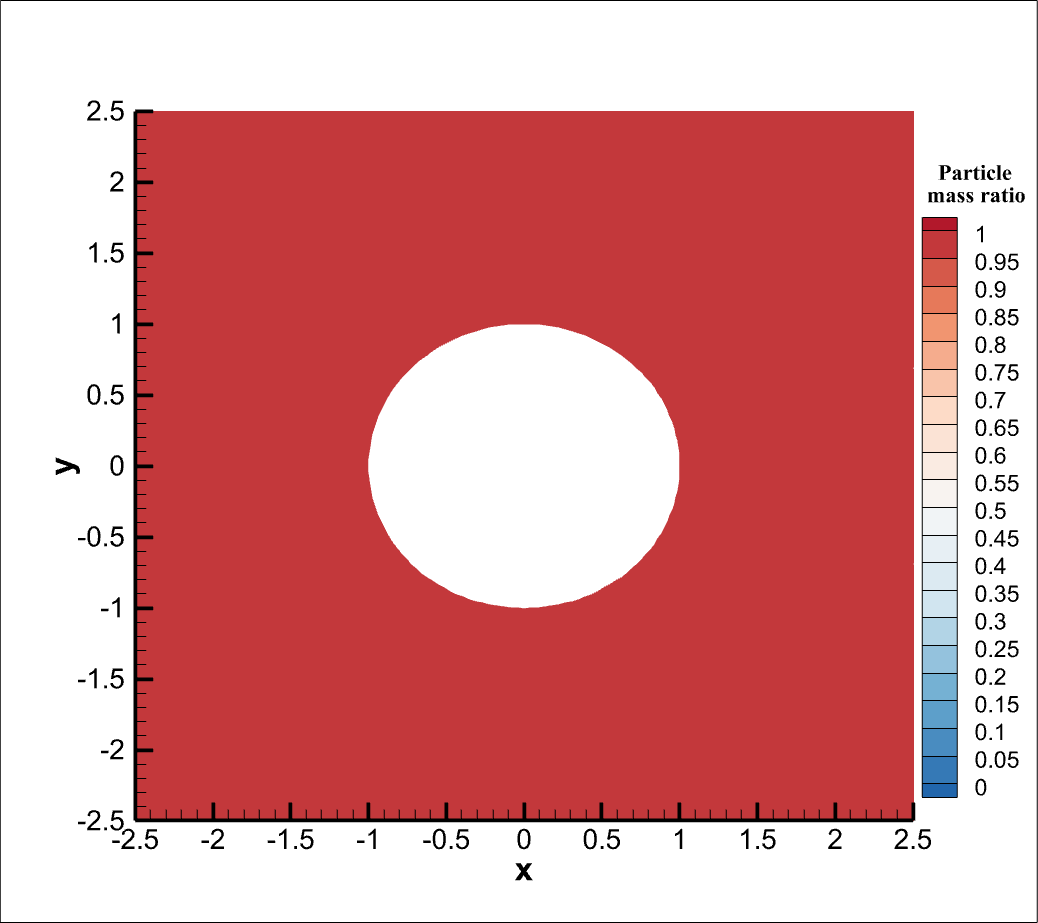}
}\hfill
\subfigure[LTS, particle mass ratio]{
\includegraphics[width=0.23\textwidth,trim=3bp 3bp 3bp 3bp,clip]{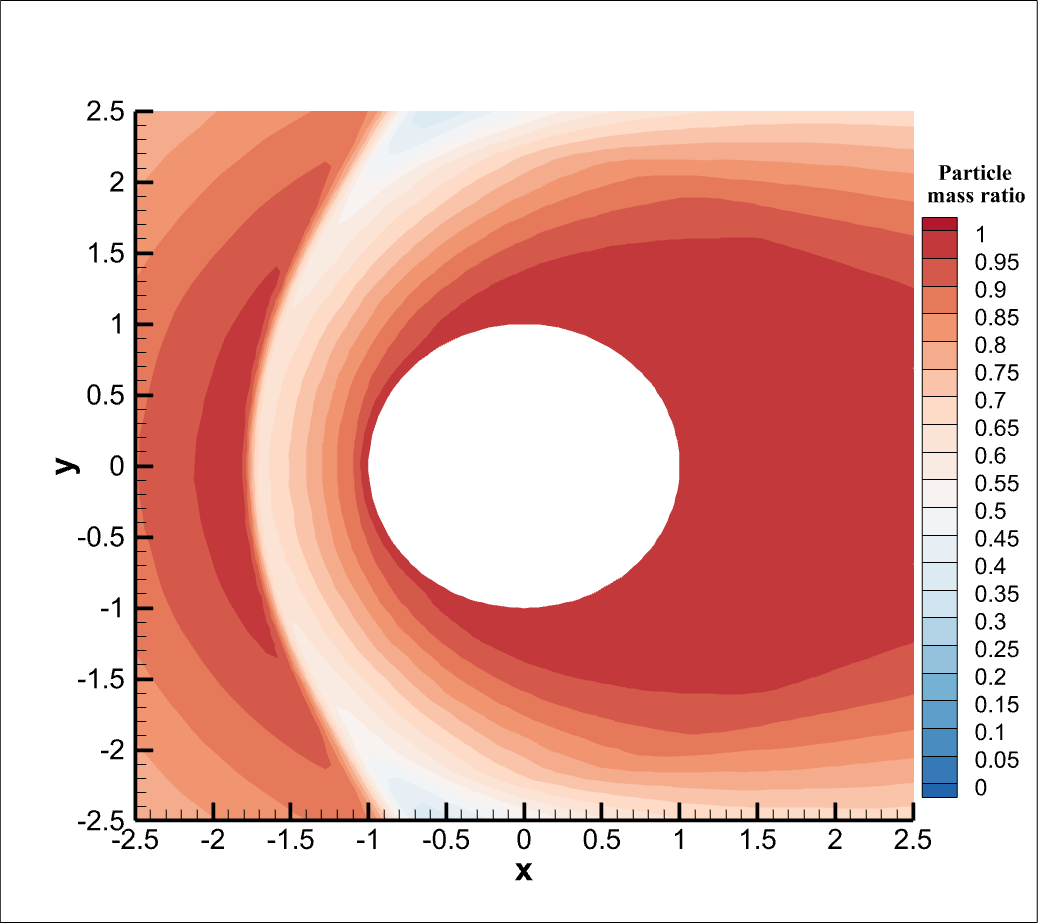}
}\\[0.5ex]
\subfigure[GTS, $\log_{10}(\Delta t_G/\tau_i)$]{
\includegraphics[width=0.23\textwidth,trim=3bp 3bp 3bp 3bp,clip]{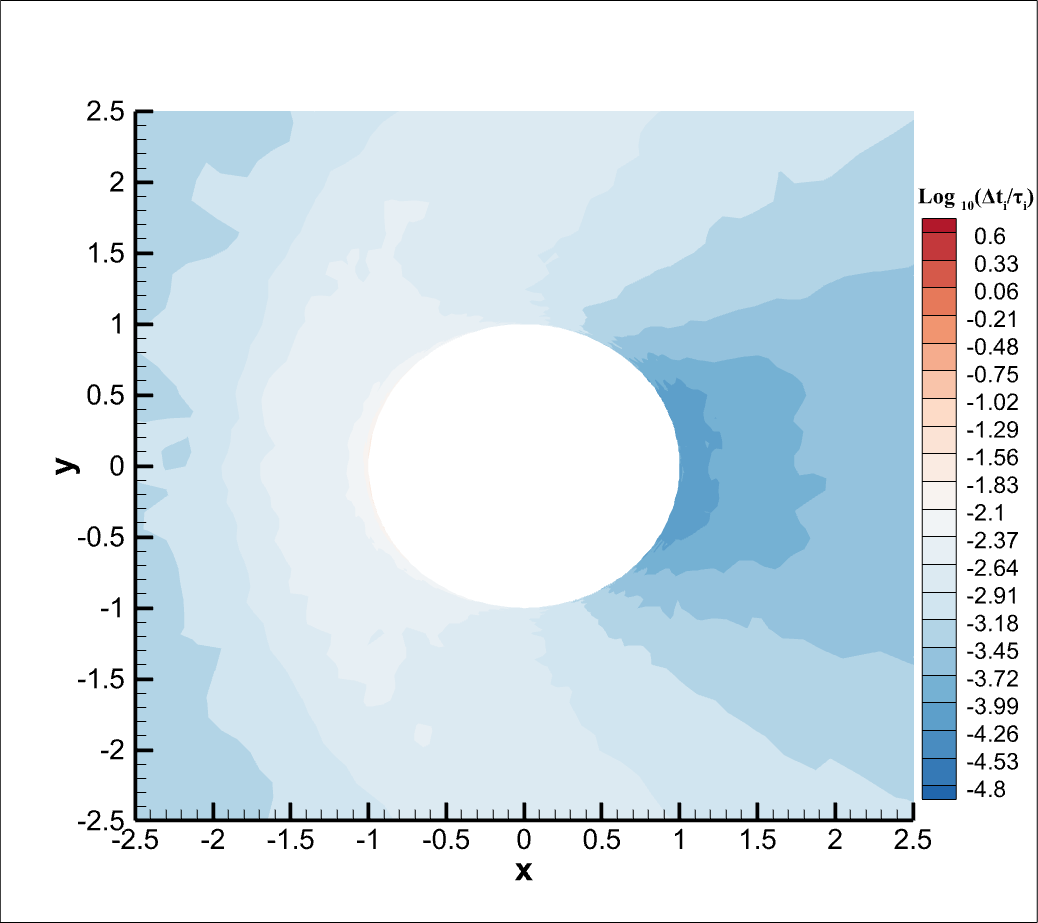}
}\hfill
\subfigure[LTS, $\log_{10}(\Delta t_i/\tau_i)$]{
\includegraphics[width=0.23\textwidth,trim=3bp 3bp 3bp 3bp,clip]{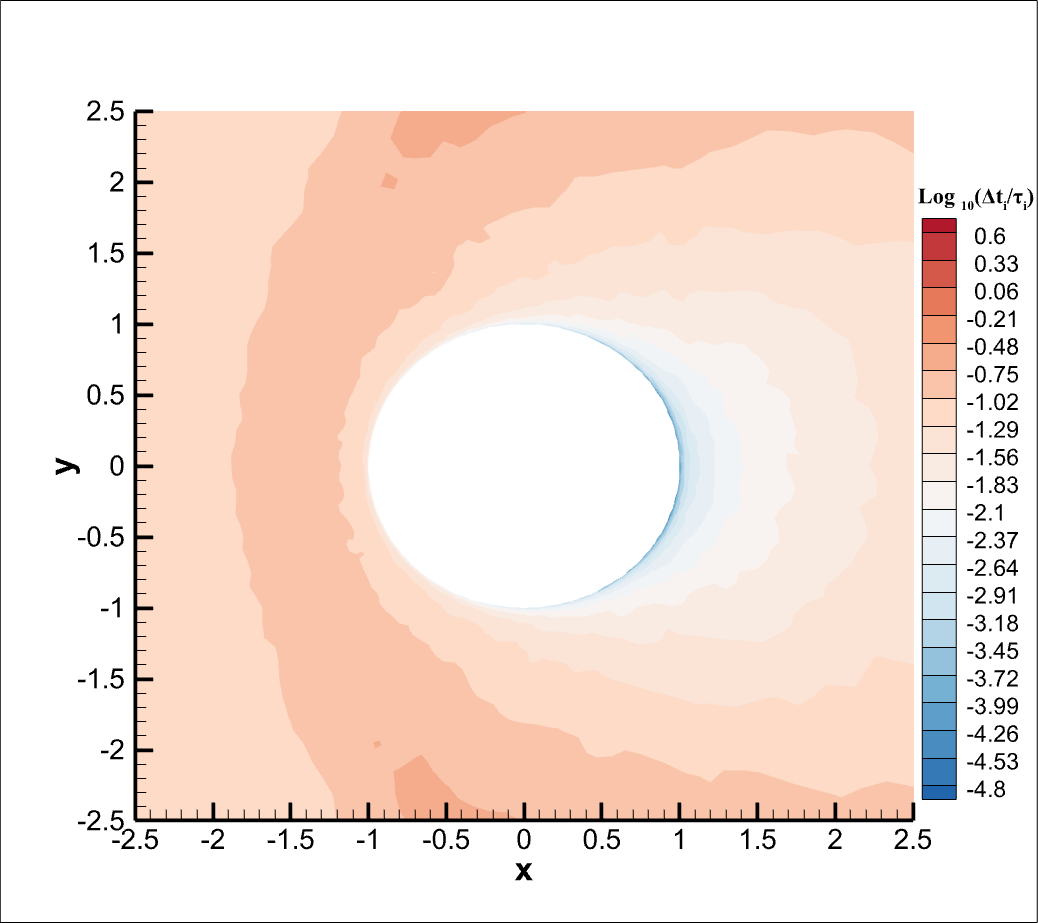}
}\hfill
\subfigure[GTS, particle mass ratio]{
\includegraphics[width=0.23\textwidth,trim=3bp 3bp 3bp 3bp,clip]{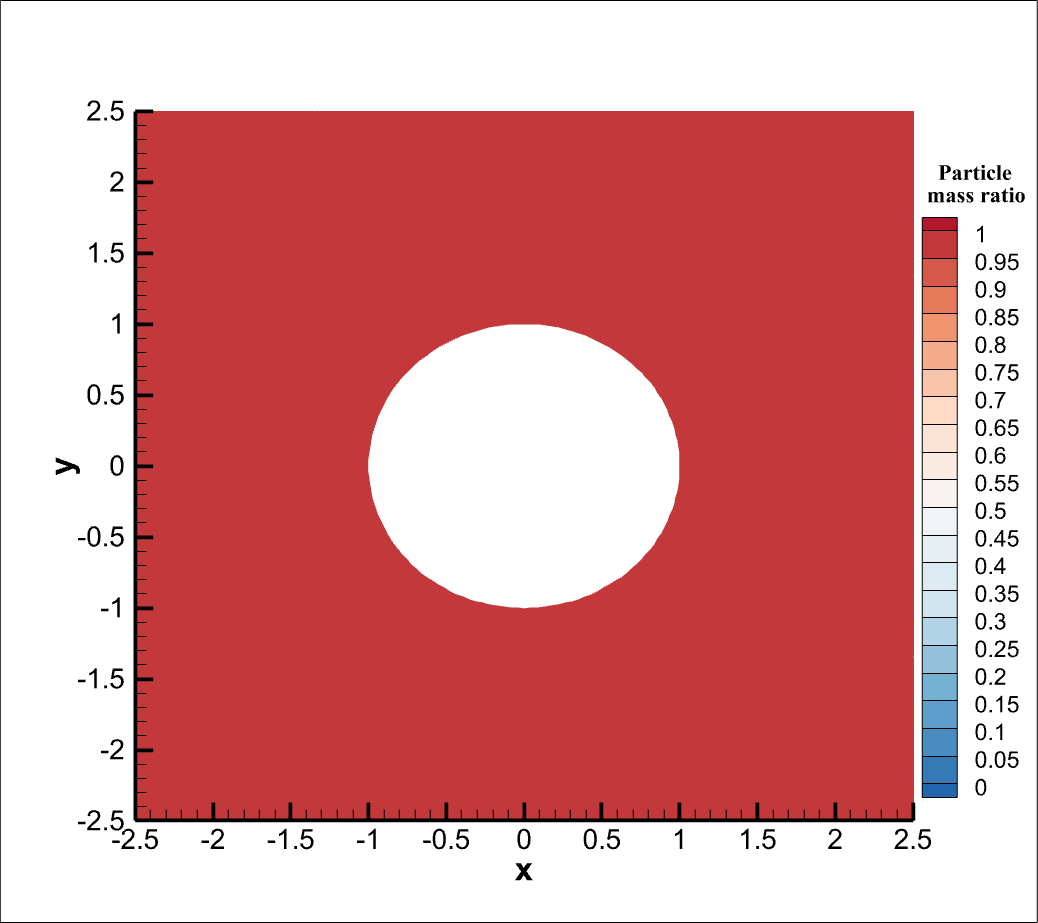}
}\hfill
\subfigure[LTS, particle mass ratio]{
\includegraphics[width=0.23\textwidth,trim=3bp 3bp 3bp 3bp,clip]{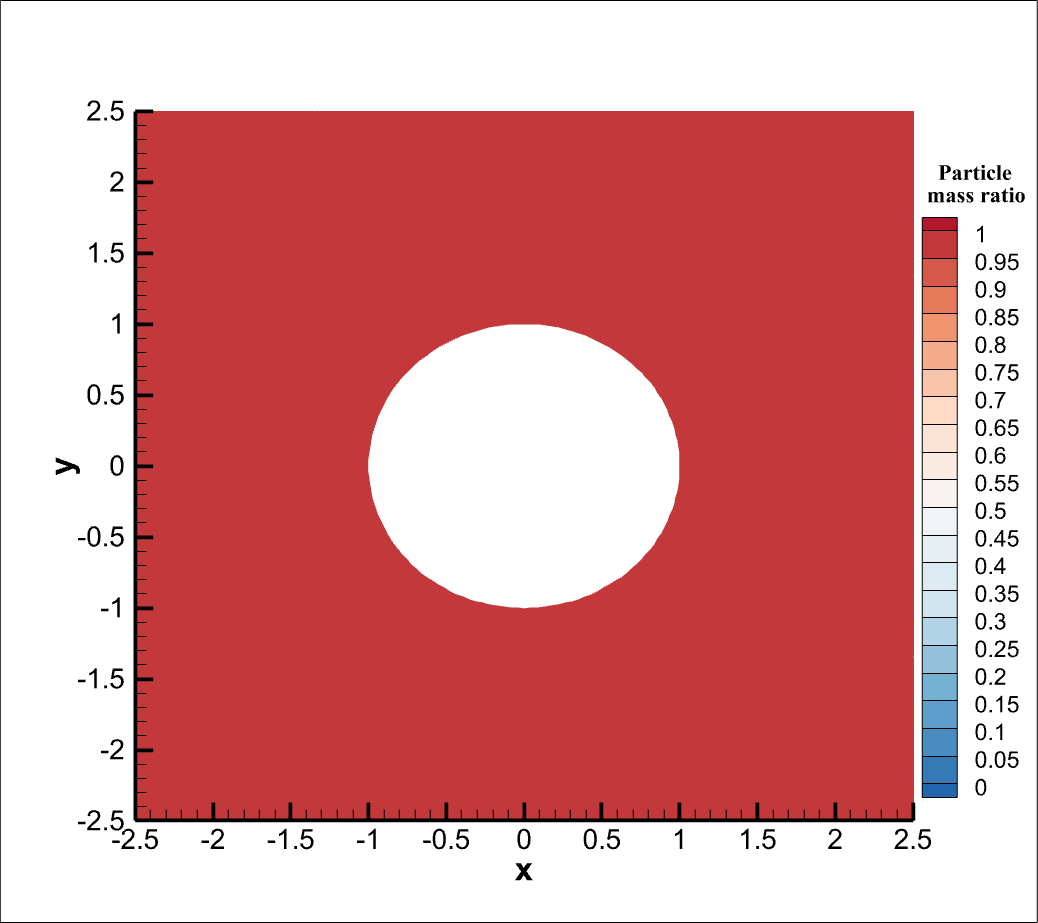}
}
\caption{\label{Fig:cyl_field}
Distributions of $\log_{10}(\Delta t/\tau_i)$ and the EMA of the particle mass ratio at step $50{,}000$ for GTS and LTS. The top and bottom rows correspond to $\mathrm{Kn}=0.01$ and $0.1$, respectively.}
\end{figure*}

Figs.~\ref{Fig:cyl_field}(a) and \ref{Fig:cyl_field}(e) show $\log_{10}(\Delta t_G/\tau_i)$ for GTS, Figs.~\ref{Fig:cyl_field}(b) and \ref{Fig:cyl_field}(f) show $\log_{10}(\Delta t_i/\tau_i)$ for LTS, and Figs.~\ref{Fig:cyl_field}(c), \ref{Fig:cyl_field}(d), \ref{Fig:cyl_field}(g), and \ref{Fig:cyl_field}(h) show the corresponding exponential moving averages (EMAs) of the particle mass ratio. The particle mass ratio in cell $i$ is defined as
\[
\begin{aligned}
  \chi_i^p &= \frac{\rho_i^p}{\rho_i},\\
  \rho_i^p &= \sum_{k\in\mathcal{N}^{+}(i)} \frac{m_k}{\Omega_i},
\end{aligned}
\]
where $\rho_i$ is the total mass density in cell $i$. The EMA of the particle mass ratio uses $\alpha=2/1001$ and is reported at step $50{,}000$. At $\mathrm{Kn}=0.01$, $\Delta t_G/\tau_i$ remains small over most of the domain in GTS, and the particle mass ratio stays close to one. LTS increases $\Delta t_i/\tau_i$ away from the near-wall region and consequently reduces the mass represented by stochastic particles, whereas the smaller $\Delta t_i/\tau_i$ near the cylinder retains a particle-dominated representation. At $\mathrm{Kn}=0.1$, the particle mass ratio remains close to one for both GTS and LTS, and the difference in wave-particle decomposition is less evident.

\begin{figure*}[!t]
\centering
\includegraphics[width=0.82\textwidth]{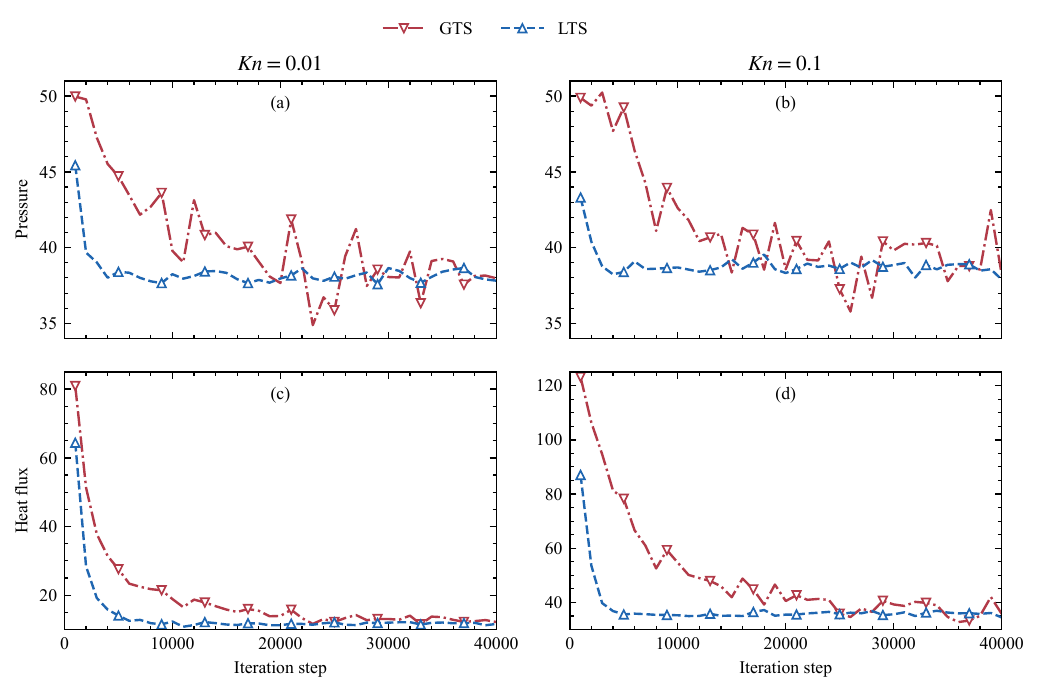}
\caption{\label{Fig:cyl_conv}
Exponential moving averages of the stagnation-point pressure and heat flux obtained with GTS and LTS at $\mathrm{Kn}=0.01$ and $0.1$, using $\alpha=2/(1000+1)$.}
\end{figure*}

The stagnation-point pressure and heat flux are obtained by linear interpolation between the two wall-face centers adjacent to the stagnation point. Figure~\ref{Fig:cyl_conv} compares the EMA histories of the stagnation-point pressure and heat flux obtained with GTS and LTS. For both cases, the LTS histories achieve steady states within approximately $5{,}000$ steps. The GTS histories achieve steady states within approximately $20{,}000$ steps and retain larger statistical fluctuations. These histories are used as qualitative indicators of convergence. The EMA smooths the statistical fluctuations in the displayed histories but introduces a lag. Therefore, the quantitative step counts reported below are determined by monitoring the stabilization of macroscopic quantities and applying practical experience, rather than directly from the EMA curves.

For the quantitative assessment at $\mathrm{Kn}=0.01$, the GTS calculation uses $20{,}000$ steps to achieve a steady state and an additional $3{,}000$ steps for averaging. The LTS calculation uses $3{,}000$ steps to achieve a steady state and an additional $500$ steps for averaging. The resulting total step counts are $23{,}000$ for GTS and $3{,}500$ for LTS. The corresponding wall-clock times are $29.9$ min for GTS and $4.2$ min for LTS. The resulting step-count and wall-clock speedups are $6.6\times$ and $7.1\times$, respectively. At $\mathrm{Kn}=0.1$, the GTS calculation uses $20{,}000$ steps to achieve a steady state and an additional $3{,}000$ steps for averaging. The LTS calculation uses $5{,}000$ steps to achieve a steady state and an additional $1{,}000$ steps for averaging. The resulting total step counts are $23{,}000$ for GTS and $6{,}000$ for LTS. The corresponding wall-clock times are $16.1$ min for GTS and $3.6$ min for LTS. The resulting step-count and wall-clock speedups are $3.8\times$ and $4.5\times$, respectively. For the $\mathrm{Kn}=0.01$ LTS calculation, the wall-clock time decreases from $48.6$ min on one core to $4.2$ min on $28$ cores, corresponding to a parallel speedup of $11.6\times$ and a parallel efficiency of approximately $41\%$.

The reductions in per-iteration computational cost and statistical fluctuations in the wall quantities are associated with the change in wave-particle decomposition produced by LTS. In GTS, the global time step is taken as the minimum of the local time steps, which generally occurs in the near-wall region. In the far-field cells, \(\Delta t_G/\tau_i\) is small, and the factor \(e^{-\Delta t_G/\tau_i}\) is close to one. A larger fraction of the hydrodynamic-wave state is therefore sampled as collisionless particles. In addition, the probability of relaxation within one global time step, $1-e^{-\Delta t_G/\tau_i}$, is small. Therefore, many of these particles remain in free transport over several steps and can reach the near-wall region. The larger mass represented by stochastic particles near the wall is associated with larger statistical fluctuations in the wall quantities. In LTS, the larger local time steps used in the far-field cells increase $\Delta t_i/\tau_i$ and reduce the fraction of the hydrodynamic-wave state sampled as collisionless particles. Consistent with the particle-mass-ratio fields, the reductions in the computational cost associated with particles and in the statistical fluctuations of the wall quantities are more evident at \(\mathrm{Kn}=0.01\) than at \(\mathrm{Kn}=0.1\).

\subsection{Hypersonic Flow Past a Flat Plate}
\label{sec:flat-plate}
The second configuration considers hypersonic flow over a flat plate~\cite{flat_Aj,flat_chen,flat_exp} of length $L=100$ mm and thickness $h=5$ mm at $\mathrm{Ma}=20.2$ and $\mathrm{Kn}=0.0169$. The reference length and temperature used for nondimensionalization are $\ell_{\mathrm{ref}}=1$ mm and $T_{\mathrm{ref}}=283$ K, respectively. The VHS model with $\omega=0.74$ is employed for nitrogen, and the reference particle number is set to $N_{\mathrm{ref}}=200$. The freestream and wall temperatures are $T_{\infty}=13.32$ K and $T_w=290$ K, respectively. A Maxwell boundary condition with an accommodation coefficient of $0.85$ is applied at the plate surface. The computational mesh contains $5{,}885$ cells, and both GTS and LTS calculations use a CFL number of $0.8$. The first-layer cell height at the plate surface is $\Delta y_{\min}=10^{-5}L$, which is one-fiftieth of the near-wall cell height used for the cylinder case at $\mathrm{Kn}=0.01$.

\begin{figure}[!tb]
\centering
\subfigure[$C_p$]{\includegraphics[width=0.72\columnwidth]{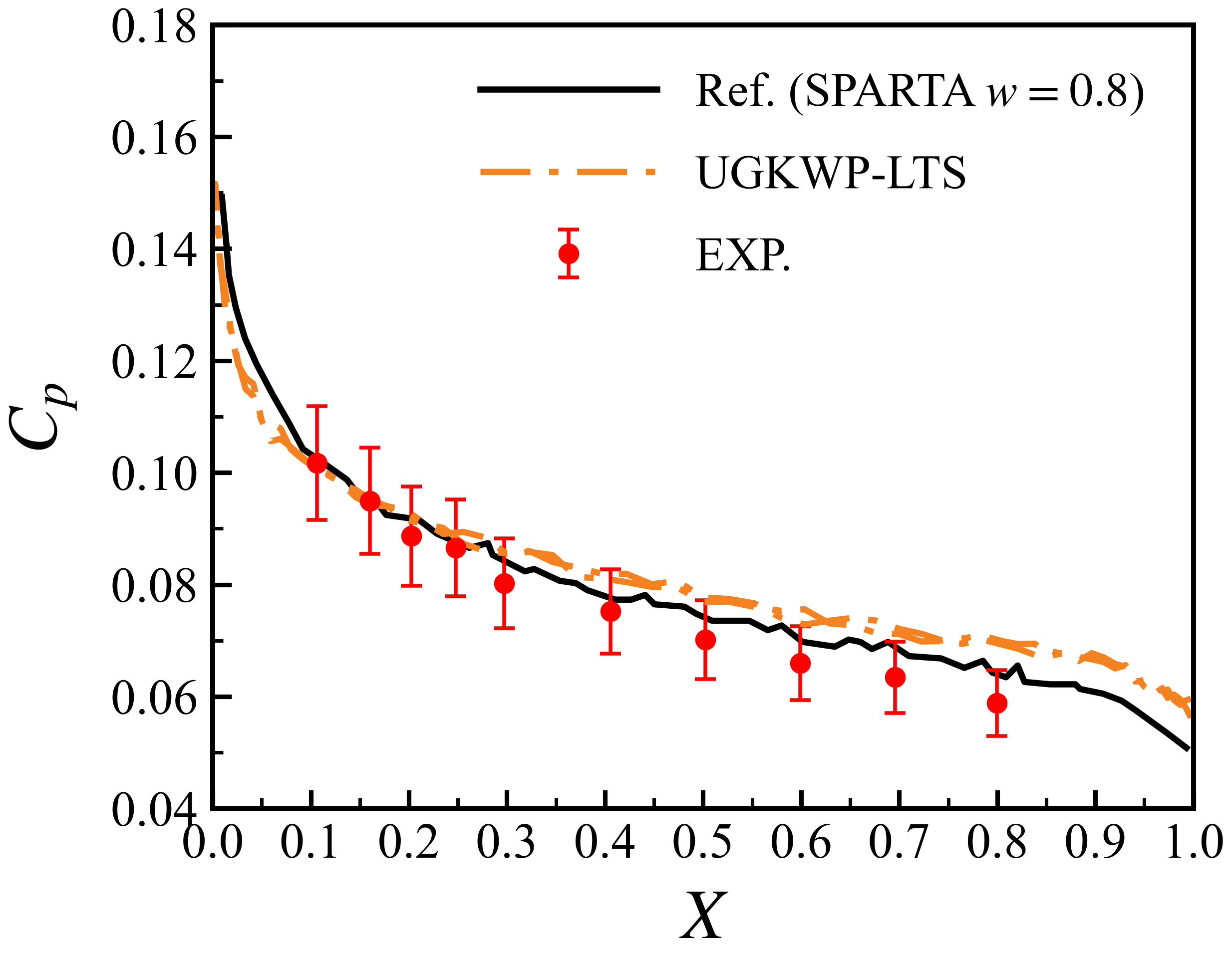}}\\[0.5ex]
\subfigure[$C_Q$]{\includegraphics[width=0.72\columnwidth]{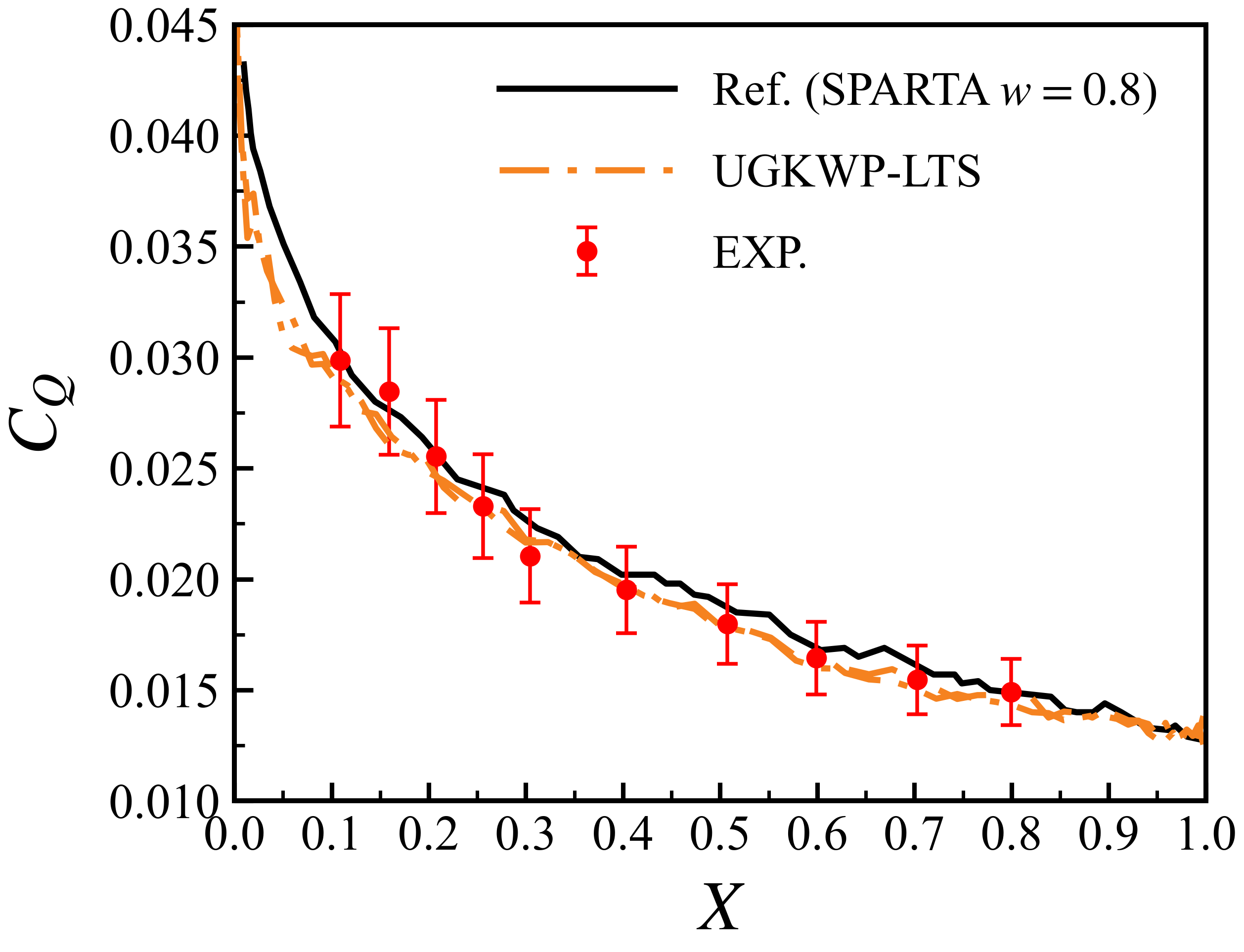}}
\caption{\label{Fig:flat_wall}Wall pressure and heat-flux coefficients along the flat plate obtained with UGKWP-LTS, compared with DSMC and experimental data.}
\end{figure}

\begin{figure}[!tb]
\centering
\includegraphics[width=0.82\columnwidth]{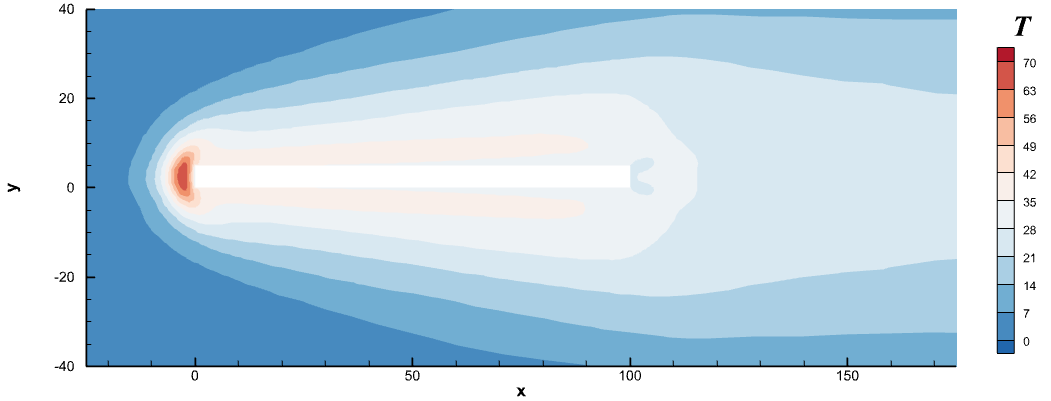}
\caption{\label{Fig:flat_field}Temperature distribution around the flat plate obtained with UGKWP-LTS.}
\end{figure}

Figure~\ref{Fig:flat_wall} compares the wall pressure and heat-flux coefficients obtained with UGKWP-LTS with the DSMC results reported with an accommodation coefficient of $0.8$ in Ref.~\cite{flat_Aj} and the experimental data in Ref.~\cite{flat_exp}. The UGKWP-LTS profiles agree closely with the DSMC and experimental data. The corresponding temperature field obtained with UGKWP-LTS is shown in Fig.~\ref{Fig:flat_field}.

\begin{figure}[!tb]
\centering
\includegraphics[width=0.82\columnwidth]{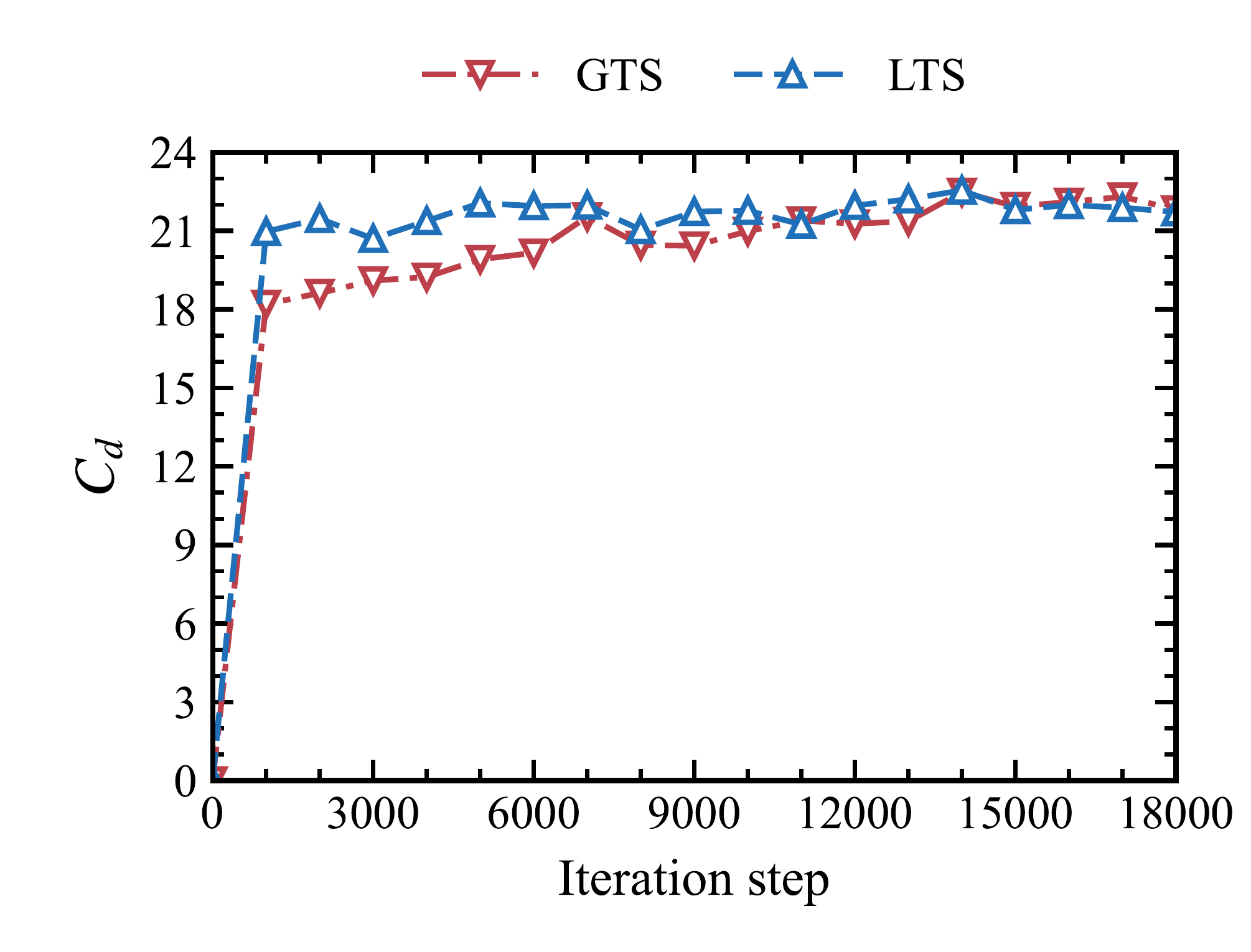}
\caption{\label{Fig:flat_conv}Convergence histories of the drag coefficient $C_d$ obtained with GTS and LTS.}
\end{figure}

The drag coefficient used to monitor convergence is defined as
\[
C_d=\frac{F_d}{\frac12\rho_\infty |\boldsymbol U_\infty|^2L},
\]
\looseness=-1
where \(F_d\) denotes the total drag. Figure~\ref{Fig:flat_conv} compares the convergence histories of the drag coefficient $C_d$ obtained with GTS and LTS.

The histories are used as qualitative indicators of convergence. The quantitative step counts below were determined by monitoring the stabilization of macroscopic quantities and applying practical experience. With GTS, the drag coefficient $C_d$ achieves a steady state after $20{,}000$ steps, followed by $50{,}000$ steps for statistical averaging, giving $70{,}000$ steps and a wall-clock time of $63$ min in total. With LTS, $C_d$ achieves a steady state after $3{,}000$ steps, followed by $500$ steps for statistical averaging, giving $3{,}500$ steps in total. The corresponding wall-clock time is estimated to be approximately $3$ min from the recorded rate of $8.5$ min per $10{,}000$ steps. These results correspond to a step-count speedup of $20\times$ and an approximate wall-clock speedup of $21\times$. The larger speedup than in the cylinder case at $\mathrm{Kn}=0.01$ is consistent with the smaller first-layer cell height, which reduces the global time step used by GTS.

\section{Conclusions}
\label{sec:conclusions}
For particle-based or hybrid wave-particle methods under FVM framework, we rigorously established that fixed positive local time steps, together with proportional particle-mass scaling, provide a sufficient condition for time-averaged interfacial particle-flux balance. Specifically, the mass of a particle crossing from cell $L$ to cell $R$ is scaled by $\Delta t_R/\Delta t_L$. We incorporated this condition into the implementation of LTS in UGKWP. The local time steps are initialized from the initial flow field according to the CFL condition and remain fixed throughout the steady flow simulation. The interfacial time-averaged wave fluxes are evaluated by integrating the equilibrium and analytic free transport wave fluxes over the corresponding cell-side local time steps and normalizing them by those time steps. The wave-particle decomposition in each cell is determined by the local ratio $\Delta t_i/\tau_i$, which better reflects the relation between the observation scale and the local relaxation time than $\Delta t_G/\tau_i$ in GTS and improves computational efficiency by increasing the wave fraction in cells where this ratio is larger. The resulting UGKWP-LTS method was evaluated for hypersonic flow past a cylinder at $\mathrm{Kn}=0.01$ and $0.1$, and for hypersonic flow past a flat plate at $\mathrm{Kn}=0.0169$. In all three cases, the surface quantities obtained with UGKWP-LTS agree closely with the reference data. Relative to GTS, UGKWP-LTS achieved step-count speedups of $6.6\times$, $3.8\times$, and $20\times$ for the three cases, respectively. The corresponding wall-clock speedups were $7.1\times$, $4.5\times$, and approximately $21\times$.

\begin{acknowledgments}
This work was supported by the National Key R\&D Program of China (Grant No. 2022YFA1004500), the National Natural Science Foundation of China (Grant No. 92371107), and the Hong Kong Research Grants Council (Grant No. 16208324).
\end{acknowledgments}

The authors have no conflicts to disclose.

\section*{DATA AVAILABILITY}
The data that support the findings of this study are available from the corresponding author upon reasonable request.

\bibliography{refs}

\end{document}